\documentclass[fleqn,usenatbib]{mnras}

\usepackage{newtxtext,newtxmath}
\usepackage{tabularx}
\usepackage[T1]{fontenc}
\DeclareRobustCommand{\VAN}[3]{#2}
\let\VANthebibliography\thebibliography
\def\thebibliography{\DeclareRobustCommand{\VAN}[3]{##3}\VANthebibliography}

\usepackage{graphicx}	
\usepackage{amsmath}	
\usepackage{relsize}
\usepackage{orcidlink}

\newcommand{\sbr}[1]{_\mathrm{#1}}
\newcommand{\sbrc}[1]{_{\mathrm{\mathsmaller{#1}}}}
\newcommand{\sigcinv}{\langle \Sigma\sbr{c}^{-1}\rangle}
\newcommand{\lcdm}{$\Lambda$CDM }
\newcommand{\ap}{Alcock-Paczy\'nski }

\newcommand{\cvk}{C_{\mathrm{V}\kappa}}
\newcommand{\ckk}{C_{\kappa \kappa}}

\title[UNIONS Void Lensing]{Lensing the darkness: The matter density profile in cosmic voids from UNIONS}

\author[H. L. Martin]{
Hunter L. Martin\orcidlink{https://orcid.org/0009-0009-1137-5880}$^{1,2}$\thanks{E-mail: h22marti@uwaterloo.ca},
Michael J. Hudson\orcidlink{https://orcid.org/0000-0002-1437-3786}$^{1,2,3}$,
Alex Woodfinden\orcidlink{https://orcid.org/0000-0002-5887-3205}$^{1,2}$,
Lucie Baumont$^{4,5}$,
\newauthor 
Thomas de Boer\orcidlink{https://orcid.org/0000-0001-5486-2747}$^{6}$,
Pierre A. Burger\orcidlink{https://orcid.org/0000-0001-8637-6305}$^{1,2}$,
Jack Elvin-Poole\orcidlink{https://orcid.org/0000-0001-5148-9203}$^{1,2}$,
S\'ebastien Fabbro\orcidlink{https://orcid.org/0000-0003-2239-7988}$^{7}$,
Samuel Farrens\orcidlink{https://orcid.org/0000-0002-9594-9387}$^{8}$,
\newauthor 
Sacha Guerrini\orcidlink{https://orcid.org/0009-0004-3655-4870}$^{8}$,
Axel Guinot\orcidlink{https://orcid.org/0000-0002-5068-7918}$^{9}$,
Fabian Hervas-Peters\orcidlink{https://orcid.org/0009-0008-1839-2969}$^{8}$,
Hendrik Hildebrandt\orcidlink{https://orcid.org/0000-0002-9814-3338}$^{10}$,
Martin Kilbinger\orcidlink{https://orcid.org/0000-0002-0805-7840}$^{8}$,
\newauthor
Magdy Morshed\orcidlink{https://orcid.org/0000-0002-3214-8881}$^{11}$
Ludovic van Waerbeke\orcidlink{https://orcid.org/0000-0002-2637-8728}$^{12}$,
Anna Wittje\orcidlink{https://orcid.org/0000-0002-8173-3438}$^{10}$
\\
$^{1}$Department of Physics and Astronomy, University of Waterloo, 200 University Avenue West, Waterloo, Ontario N2L 3G1, Canada\\
$^{2}$Waterloo Centre for Astrophysics, University of Waterloo, Waterloo, Ontario N2L 3G1, Canada\\
$^{3}$Perimeter Institute for Theoretical Physics, 31 Caroline St. North, Waterloo, ON N2L 2Y5, Canada\\
$^{4}$Dipartimento di Fisica - Sezione di Astronomia, Università di Trieste, Via Tiepolo 11, 34131 Trieste, Italy \\
$^{5}$INAF-Osservatorio Astronomico di Trieste, Via G. B. Tiepolo 11, 34143 Trieste, Italy\\
$^{6}$Institute for Astronomy, University of Hawaii, 2680 Woodlawn Drive, Honolulu HI 96822, USA\\
$^{7}$NRC Herzberg Astronomy \& Astrophysics, 5071 West Saanich Road, British Columbia V9E2E7, Canada
\\
$^{8}$Universit\'e Paris Cit\'e, Universit\'e Paris-Saclay, CEA, CNRS, AIM, 91191, Gif-sur-Yvette, France\\
$^{9}$Department of Physics, McWilliams Center for Cosmology, Carnegie Mellon University, Pittsburgh, PA 15213, USA\\
$^{10}$Ruhr University Bochum, Faculty of Physics and Astronomy, Astronomical Institute (AIRUB), German Centre for Cosmological Lensing, 44780 Bochum, Germany\\
$^{11}$INFN Sezione di Ferrara, Via Saragat 1, 44122 Ferrara, Italy\\
$^{12}$Department of Physics and Astronomy, University of British Columbia, 6224 Agricultural Road, V6T 1Z1, Vancouver, Canada\\
}

\date{Accepted XXX. Received YYY; in original form ZZZ}

\pubyear{2025}

\begin{document}
\label{firstpage}
\pagerange{\pageref{firstpage}--\pageref{lastpage}}
\maketitle

\begin{abstract}
We measure the distribution of matter contained within the emptiest regions of the Universe: cosmic voids. We use the large overlap between the Ultraviolet Near-Infrared Optical Northern Survey (UNIONS) and voids identified in the LOWZ and CMASS catalogues of the Baryon Oscillation Spectroscopic Survey (BOSS) to constrain the excess surface mass density of voids using weak lensing. We present and validate a novel method for computing the Gaussian component of the conventional weak lensing covariance, adapted for use with void studies. We detect the stacked weak lensing void density profile at the $6.2\sigma$ level, the most significant detection of void lensing from spectroscopically-identified voids to date. We find that large and small voids have different matter density profiles, as expected from numerical studies of void profiles. This difference is significant at the $2.3\sigma$ level. Comparing the void profile to a measurement of the void-galaxy cross-correlation to test the linearity of the relationship between mass and light, we find good visual agreement between the two, and a galaxy bias factor of $2.45\pm0.36$, consistent with other works. This work represents a promising detection of the lensing effect from underdensities, with the goal of promoting its development into a competitive cosmological probe.
\end{abstract}



\section{Introduction}
\label{sc:Intro} 
The cosmic web is a network of filaments and clusters that developed as the Universe evolved. Spanning the empty space between the filaments are cosmic voids: large underdense regions that formed as matter from the homogeneous early-time Universe was drawn into the overdense structures that comprise the cosmic web. These cavities provide an area that is low density, insensitive to baryonic effects, and subject to primarily linear physics down to the smallest scales \citep{Schuster2023, Schuster2024}. This creates an environment that can test established physics in low-density extremes. As an example, because of the relative absence of dark matter, void physics is relatively more sensitive to dark energy physics. Further, void sizes are on the typical order of neutrino free-streaming scales, meaning that voids are expected to carry neutrino mass signatures \citep{Zhang2020}. Additionally, as voids form via gravitational effects, modified gravity models like fifth-force models typically produce emptier voids due to the increased gravitational strength \citep{Cai2015,Cautun2018,Paillas2019,Davies19}. This has led to voids rapidly developing as a promising source of cosmological information over the course of the past decade. To probe that information, it has been demonstrated that void statistics like the void size function \citep{SvdW04,Jennings13,Pisani15,Contarini2023,FernandezGarcia2025}, and the void-galaxy cross-correlations \citep{Hamaus15,Hamaus2016,Nadathur2020b, Hamaus2020,Woodfinden2022,Woodfinden2023,Fraser2024}, are sensitive to cosmology. Voids can also be combined with galaxy statistics to break known degeneracies and offer complementary information, leading to overall tighter constraints \citep{Bayer2021,Kreisch22,Pelliciari2023,Thiele2024,Contarini2024,Salcedo25}. Additionally, work on simulated voids has revealed that the matter density profile of voids is particularly sensitive to modified gravity models \citep{Cai2015,Barreira2015,Paillas2019,Davies19} and neutrino mass \citep{Massara2015,Kreisch2019,Schuster19,Vielzeuf2023}. The primary method of measuring this latter statistic is through weak gravitational lensing.  

In the traditional weak lensing scenario, the gravitational potential from an overdensity distorts the image of background sources of light, creating tangential alignment of the images around the foreground mass. We can treat voids as having a negative overdensity, flipping the direction of the distortion, resulting in radial alignment. However, galaxies have overdensities of more than $200$ times the density of the Universe, while voids cannot be less dense than a pure vacuum, having an underdensity of $-1$. Voids, therefore, have a significantly weaker lensing effect on background sources than do virialized overdensities. Consequently, the detection of weak lensing by voids is observationally challenging. 

The first detection of a radial weak shear signal from underdensities identified using photometric redshifts was obtained by \cite{Gillis13}, and the first detection from spectroscopically-identified voids by \cite{Melchior2014}, followed by subsequent measurements primarily using photometric redshifts \citep{Clampitt15,Sanchez16,Fang2019,Jeffrey2021}. Additionally, voids can also be used to study lensing signals in the Cosmic Microwave Background (CMB) \citep{Cai2017,Vielzeuf2023,Camacho-Ciurana2024,Demirborzan2024}.

One of the reasons this void lensing detection is challenging comes from the definition of the voids themselves. Because the underlying matter density is not directly known, and because voids do not provide any direct observational evidence for their location, there is ambiguity in defining the location and boundaries of a void. As a result, various methods of finding voids are described in the literature. For example, 2D void finders, such as those used in \citet{Gruen16} and \citet{Fang2019}, identify voids as circular underdensities in projected tracer fields, while void finders based on the ZOnes Bordering on Voidness (ZOBOV) algorithm \citep{Neyrinck08}, such as those used in \citet{Melchior2014} and \citet{Hamaus2014}, identify 3D underdensities in the unprojected tracer field. As shown in various comparative analyses \citep{Colberg2008,Cautun2018,Paillas2019,Massara2022}, different void finders produce void statistics with different attributes. One such example of this is the notion that 2D voids tend to exhibit stronger projected density contrasts in their matter density profile than ZOBOV voids, as ZOBOV voids are necessarily surrounded by an overdense wall that intersects the line of sight both in front of and behind the void. This property need not exist for underdensities identified in projection. However, ZOBOV voids maintain spherical symmetry in their stacked profiles, allowing for more direct modelling and interpretation. Therefore, void analyses have the freedom of choosing void finders that are best suited to the situation. In order to utilize the overlap between the Baryon Oscillation Spectroscopic Survey (BOSS) spectroscopic survey and the Ultraviolet Near-Infrared Optical Northern Survey (UNIONS), we are interested in studying the weak lensing signal of 3D spectroscopic voids.

Additionally, properties of void statistics heavily depend on properties of the tracer field in which they are found. These include tracer bias, tracer sparsity, completeness, and availability of photometric or spectroscopic redshifts \citep{Sutter14,Nadathur2015,Massara2022,Schuster2023}. As an example, \citet{Fang2019} has found that ZOBOV voids identified in photometric galaxy surveys exhibit a selection effect that extends the average void shape along the line of sight, producing voids with deeper density contrasts than expected for more spherical voids. In the particular case of tracer bias, \citet{Pollina17} demonstrated through comparisons in hydrodynamical simulations of void tracer profiles and void matter profiles that a linear biasing model is sufficient to relate the void-galaxy profile to the void-matter profile, but that the bias changes with void radius. They found that the void-galaxy bias (what we will call $b\sbr{Vg}$) is larger than the large-scale linear galaxy bias ($b\sbr{g}$) for small voids, then asymptotically approaches $b\sbr{g}$ with increasing void size. This effect was tested using DES Y1 data in \citet{Fang2019}, who found consistency with the large-scale bias irrespective of void size, but with large enough scatter to mask any trend. In contrast, \citet{Nadathur2019b}, using $N$-body simulations with Halo Occupation Distribution (HOD) modelling, argue from a stochastic basis that clustering of a tracer sample around voids identified within that tracer sample produce a selection effect. This then biases the stacked void-galaxy profile such that the galaxy bias depends on the distance from the centre of the void, with a fractional difference up to 25 per cent. One of the goals of this paper is therefore to test the linear bias model using voids identified in spectroscopic data by comparing our void lensing measurement to the void-galaxy correlation measurement from \citet{Woodfinden2022}.

The aim of this paper is to measure and determine the significance of the void lensing signal of 3D spectroscopic voids coming from current generation surveys with the intention to demonstrate the potential for next-generation surveys such as \textit{Euclid} \citep{Euclid2025} and Roman \citep{Roman2021}, both of which promise larger and higher-quality data sets than currently available. We are additionally interested in how well we can constrain the large-scale structure information contained within this signal. As such, the paper is organized as follows. Section~\ref{sc:data} describes the data catalogues, including the UNIONS sources and BOSS voids found by \citet{Woodfinden2022}. In Section~\ref{sc:Methods}, we detail the methodology for obtaining the lensing measurement for these voids. This includes methodology for obtaining the lensing signal around random voids and the novel derivation of an analytic covariance model. Additionally, we describe both models used in this analysis: an analytic model used to assess the significance of the detection, and a void-galaxy measurement to constrain linear void-galaxy bias. In Section~\ref{sc:results}, we present our measurements. In Section~\ref{sc:discussion_header}, we compare and discuss our measurements in comparison with other works and discuss future work in light of upcoming galaxy imaging and clustering surveys. The fiducial cosmology we assume for this work is a flat \lcdm cosmology with $\Omega\sbr{m}$ = 0.307, $\Omega\sbr{b}$ = 0.0482, $h$ = 0.6777, $\sigma_8$ = 0.8225, and $n\sbr{s}$ = 0.96, and all distances are in comoving units.

\section{Data}
\label{sc:data}
\subsection{UNIONS source galaxies}
\label{sc:UNIONS Sources}
\begin{figure}
    \centering
    \includegraphics[width=\columnwidth]{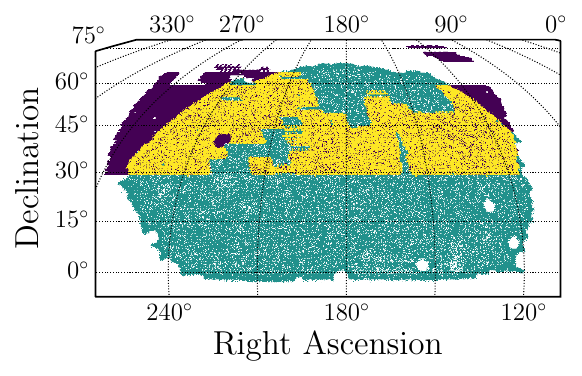}
    \caption{Footprints of the two surveys used in this analysis and their overlapping region. The UNIONS footprint at the time of writing is in purple, and the CMASS footprint is shown in light blue. The overlapping region of the two surveys is shown in yellow.}
    \label{fig:UNIONS}
\end{figure}

The catalogue of background source galaxy shapes that are used in this analysis are from the Ultraviolet Near-Infrared Optical Northern Survey (UNIONS) \citep{Gwyn2025}. UNIONS is a planned 6250 deg$^2$ multi-band survey in the northern hemisphere, with about 3200 deg$^2$ completed for use in this work. This is the largest lensing survey of the northern hemisphere to date, and offers significant overlap with spectroscopic surveys like SDSS. This overlap makes UNIONS an ideal choice for this work. UNIONS is comprised of the CFIS, Pan-STARRS, and WISHES surveys, each providing data in different broad band filters. In the future, these data sets will be combined for full \textit{ugriz} photometry, but individual photometric redshifts and lensing tomography are in preparation and therefore currently unavailable for this work. The source galaxy shapes are measured using the $r$-band data provided by CFIS using the Canada-France Hawai'i Telescope (CFHT) on Mauna Kea. The $r$-band data is planned to reach a depth of $r=24.2$ at $\mathrm{SNR}=10$ with $\approx 0.7$ arcsec seeing \citep{Gwyn2025}. This is ideal for measuring galaxy shapes for weak lensing measurements. The footprint of the UNIONS survey showing the overlap with the foreground SDSS BOSS CMASS survey is shown in Fig.~\ref{fig:UNIONS}

As described in \citet{Li2024}, in order to build the redshift distribution, the UNIONS survey uses the overlap between the UNIONS $r$-band footprint and the CFHTLS-W3 field \citep{Heymans2012}. UNIONS galaxies in the overlap are then matched to their corresponding galaxy image in the public W3 field to assign full $ugriz$ photometry found in the Canada-France-Hawaii Telescope Lensing Survey (CFHTLenS) \citep{Hildebrandt12}. This is then used to populate a self-organizing map (SOM) trained on publicly available overlapping spectroscopic surveys \citep{Wright20}. The spectroscopic redshift distribution is then reweighted by the ratio of UNIONS galaxies to spectroscopic galaxies in each SOM cell, resulting in a calibrated source redshift distribution. Then, given that UNIONS has a constant depth across the survey area, we are justified in using this redshift distribution as representative across the entire survey. This distribution is shown along with the redshift distribution of foreground lenses in Fig.~\ref{fig:nvz}.

The shapes of these source galaxies were measured using version 1.3 of the \texttt{ShapePipe} pipeline, the details of which are outlined in \citet{Guinot22} and \citet{Farrens2022}. Version 1.3 of the pipeline uses the \texttt{MCCD} \citep{Liaudat2021} software package to model the PSF \citep{Li2024,Guerrini2025,Zhang2024}, and continues to use \texttt{NGMIX} to perform \texttt{METACALIBRATION} \citep{Metacal17}. The result is a catalogue that includes the sky position, the two ellipticity components $\epsilon_1$ and $\epsilon_2$, as well as an inverse variance weight that accounts for the quality of the shape measurement:
\begin{equation}
    \label{eq:sourceweight}
    w\sbrc{S}=\frac{1}{2\sigma_\epsilon^2+\sigma^2_{\epsilon_1}+\sigma^2_{\epsilon_2}}\,,
\end{equation}
where $\sigma_{\epsilon}$ represents the intrinsic dispersion of one component of the ellipticity, set to the same value of 0.34 as used in \citet{Guinot22}, and the other two terms are the measurement errors in each corresponding ellipticity component from a model fit. The 1.3 version of this catalog includes $83\,812\,739$ total galaxy shapes and represents the status of the data as of 2022.
\subsection{Foreground Voids}
\label{sc:Voids Header}
\subsubsection{BOSS Survey}
\label{sc:BOSS}
The catalogue of lenses that are used in this analysis comes from the Baryon Oscillation Spectroscopic Survey (BOSS) \citep{Dawson13} void catalogue of \cite{Woodfinden2022}. The final BOSS survey is the 12th Data Release of the Sloan Digital Sky Survey (SDSS-III) \citep{Eisenstein11}, a spectroscopic survey targeting luminous red galaxies (LRGs) with two different targeting algorithms, LOWZ and CMASS \citep{Reid16}. LOWZ targeted luminous red galaxies (LRGs) in the nearby universe ($0.2<z<0.43$) to produce a volume-limited survey. CMASS targeted LRGs just beyond the LOWZ sample ($0.43<z<0.75$) with the goal of targeting galaxies with constant stellar mass over the redshift range. Both surveys used the 2.5m Sloan telescope at the Apache Point Observatory, and covered roughly the same area, with LOWZ being smaller than CMASS. The total survey area is around 10,000 $\textrm{deg}^2$. To increase the statistical power of our measurement, we can combine voids found in both the LOWZ and CMASS LRG catalogues. This is reasonable because the galaxy clustering amplitudes of the two LRG samples are within 20 per cent of each other \citep{Ross17}, meaning that the tracer populations used to identify the voids will exhibit similar properties, including similar galaxy bias values.

\subsection{Void Finding}
\label{sc:Void Find Header}
We will now summarize the steps of the void finding implementation. For more specific details, we refer readers to \cite{Woodfinden2022} for the implementation that was applied to our data set, and the references therein for more details on the void finding code \texttt{Revolver} \citep{Nadathur2019a}. 

\subsubsection{Reconstruction}
\label{sc:Reconstruct}
Individual voids can be highly non-spherical, but stacked voids under the assumption of cosmological isotropy can be used as a standard spherical shape in cosmology, and are a candidate for the \ap (AP) test. However, redshift-space distortions (RSD), caused primarily by the peculiar velocities of tracer galaxies along the line of sight, can artificially shift redshift estimations when compared to a tracer's real-space position. Typically, void finding is done in redshift space, where observations are made, resulting in an artificial elongation along the line of sight, even for spectroscopic tracers where redshifts are well-estimated. To correct for this effect, the \texttt{Revolver} void finder as described in \citet{Nadathur2019a} first attempts to reconstruct the real-space positions of tracers before running the void finder. RSDs are removed by solving the Zel'dovich equation in redshift space for the displacement field $\mathbf{\Psi}$:
\begin{equation}
    \nabla \cdot \mathbf{\Psi}+\frac{f}{b}\nabla\cdot\left(\mathbf{\Psi}\cdot\boldsymbol{\hat{\mathbf{r}}}\right)\,\boldsymbol{\hat{\mathbf{r}}}=\frac{\delta\sbr{g}}{b}\,,
\end{equation}
then galaxies are shifted to their real-space position using only the component of $\mathbf{\Psi}$ that contributes directly to RSD, $-\mathbf{\Psi}\sbr{RSD}$, defined as
\begin{equation}
    \mathbf{\Psi}\sbr{RSD}=-f\left(\mathbf{\Psi}\cdot\boldsymbol{\hat{\mathbf{r}}}\right)\,\boldsymbol{\hat{\mathbf{r}}}\,.
\end{equation}
This procedure introduces a dependence on $\beta \equiv f/b$ for our void catalogues, where $f$ is the linear growth factor, and $b$ is the linear galaxy bias. For the data set provided, $\beta$ was set to 0.37, chosen to be the closest match to the fiducial cosmology.

\subsubsection{Void Finding Procedure}
\label{sc:Void_finder}
Once the set of tracer galaxies has been reconstructed to their real-space positions, they are passed to a modified ZOBOV void-finder \citep{Neyrinck08} provided by \texttt{Revolver}. ZOBOV algorithms start with a Voronoi tessellation of the galaxy field, where each cell includes a single galaxy and all spatial points that are closer to that galaxy than any other galaxy in the sample. This tessellation means that cells in high-density regions (i.e.\ many tracer points) will have predominantly small volumes, while low-density regions will have predominantly large volumes. Cells are then merged into zones using a watershed algorithm to identify the extent of local density minima of the tessellated field. Each zone is then defined as a void. 

To modify ZOBOV for use outside of simulated data, it has to account for observational effects such as survey masks. Around the edges of the survey mask, voids can ``leak'' past observed data into arbitrarily low-density regions, resulting in a subsample of voids that do not reflect properties of the underlying matter density. In order to correct these problems, a set of barrier particles is added to the galaxy field, and any Voronoi cell or zone that touches a barrier particle is removed from consideration. These create a selection bias towards smaller voids near the mask edges, necessitating careful consideration when constructing catalogues of randomly positioned lenses for the void lensing measurement as described in Sect.~\ref{sc:randcat}. 

Voids are identified as the full extent of a watershed basin around a local minima in the Voronoi tessellation. These basins are usually not spherical, so the definition of the void centre is not unique. By default in the \texttt{Revolver} code, the void centre is the circumcentre, defined to be the centre of the empty sphere that can be inscribed within the basin. The other commonly used option would be a volume-weighted barycentre. This choice has significant effects on the resulting matter profile \citep{Nadathur2015b,Massara2022}. By construction, a circumcentre definition identifies the minimum density point of the void to be the centre, whereas the barycentre need not coincide with the minimum density point. However, the definition of the barycentre is sensitive to the non-sphericity of the void, whereas the circumcentre, by using a circumscribed sphere, is not. This results in barycentre voids showing more prominent overdense walls with shallower central underdensities in a stacked profile, while circumcentre voids will have deeper central underdensities but less prominent walls \citep{Nadathur2015b}. Therefore, there is no accepted standard for the definition of these centres, but choosing a definition changes the properties of the resulting profiles. The volume of the void is taken to be the volume of all component Voronoi cells of the basin. The void radius is defined to be the radius of a sphere of equivalent volume
\begin{equation}
    R\sbr{V}= \left( \frac{3}{4\pi} V\sbr{basin}\right)^{1/3}\,.
\end{equation}

\subsubsection{Application to the BOSS Void Catalogues}
\label{sc:Void Catalog}
As discussed in \citet{Woodfinden2022}, for LOWZ and CMASS, the reconstruction step was applied to the combined sample of \cite{Alam17}, but the ZOBOV step was applied to each catalogue separately, due to differences in the survey masks. This was achieved by first adding a layer of barrier particles at z=0.43, the redshift boundary between the two surveys. This means that no voids are allowed to span between the LOWZ and CMASS surveys. The resulting LOWZ and CMASS void catalogues contain 2147 and 4781 voids in the North Galactic Cap (NGC), respectively. This total sample of voids will be used to measure the void-galaxy cross-correlation as discussed in Sect.~\ref{sc:void-gal measure}.

It should be noted that not all of the CMASS LRGs were used in this void catalogue. Due to a partial overlap of the eBOSS+CMASS void cosmology analysis of \citet{Nadathur2020b}, LRGs at $z>0.63$ were excluded from the void finding step as outlined in \citet{Woodfinden2022}. Further, to avoid issues with edge effects in the void finding step, only voids with centres at $z<0.6$ are included in the data set.

For void lensing, we are primarily interested in the subset of voids that have significant overlap with the UNIONS footprint, as these voids will be guaranteed to have source galaxies for the weak lensing calculation. In order to perform this cut, we use the $\texttt{HEALPIX}$ \citep{Gorski2005} mask of the UNIONS survey with $N_{\textrm{side}}=1024$, and we build a $\texttt{HEALPIX}$ disk around each void centre with radius equal to the radius of the void.

\begin{figure}
    \centering
    \includegraphics[width=\columnwidth]{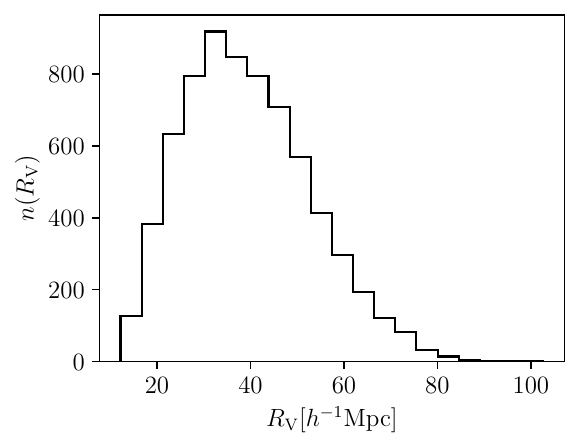}
    \caption{Void size distribution for the BOSS LOWZ+CMASS void catalogue from \citet{Woodfinden2022}. }
    \label{fig:nrv}
\end{figure}

\begin{figure}
    \centering
    \includegraphics[width=\columnwidth]{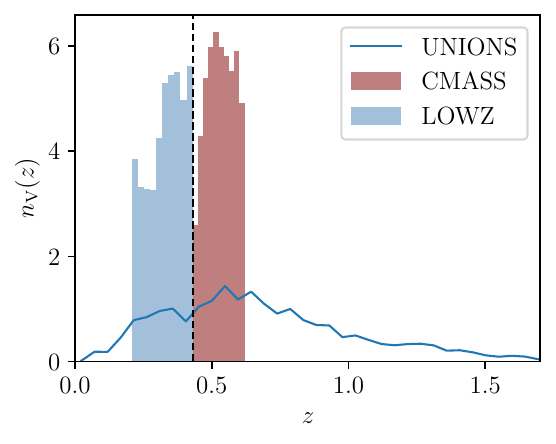}
    \caption{Void redshift distribution for the BOSS LOWZ and CMASS void catalogues from \citet{Woodfinden2022} with the UNIONS source redshift distribution. The UNIONS source distribution plotted in this figure is one of the blinded distributions. This is done in order to preserve the blinding of the cosmological analysis. It is close, but not identical, to the unblinded redshift distribution which is used for the analysis in this paper.}
    \label{fig:nvz}
\end{figure}

Comparing the disk with the mask, we include the void in our lens sample if there are any pixels that overlap. Given that we are binning out to three times the void radius in our measurement, this condition means that we exclude most of the voids that would contribute at most a small number of lens-source pairs. After this cut, we are left with 944 and 2031 voids for LOWZ and CMASS void lensing, respectively. The radius and redshift distribution are shown in Figs.~\ref{fig:nrv} and \ref{fig:nvz}, respectively. We will measure the void lensing signal coming from each catalogue individually, as well as the combination of the two, which we will label Full.

We also construct two additional catalogues by splitting the Full catalogue into large and small voids. In order to split the lensing information evenly, this split is done using the average void radius, but rather than a simple average, it is weighted by the lensing weights, which will be described below in Sect.~\ref{sc:weaklens}, summed over all lens-source pairs out to $3R\sbr{V}$. The weighted average void radius used to perform this split is $48.5 h^{-1}$ Mpc. We note that this weighting upweights larger voids as they overlap more source galaxies than smaller voids. These void size catalogues will allow us to examine the void size evolution in the matter profiles. Additionally, we can observe how the galaxy bias evolves with void size to compare to the trend observed in simulations \citet{Pollina17}. All five void catalogues and their statistics, as well as the catalogue used in the void-galaxy cross-correlation measurement, are compiled in Table~\ref{tab:void_cats}. 
\begin{table}
    \centering
    \caption{Total number, average unweighted void radius, and average redshift of the 5 void catalogues used in the void lensing analysis, as well as for the void sample used in the full void-galaxy cross-correlation measurements.}
        \begin{tabular}{l|r|c|c}
            \hline
            Catalogue & $N\sbr{Void}$ & $\bar{r}\sbr{V} [h^{-1} \textrm{Mpc}]$ & $\bar{z}\sbr{V}$ \\
            \hline
            Full & 2975 & 40.3 & 0.467 \\
            LOWZ & 944 & 40.3 & 0.331 \\
            CMASS & 2031 & 40.3 & 0.530 \\
            Small & 2167 & 33.7 & 0.466 \\
            Large & 808 & 57.9 & 0.469 \\
            Void-Galaxy & 6928 & 39.5 & 0.468
        \end{tabular}
    
    \label{tab:void_cats}
\end{table}

\subsubsection{\texttt{PATCHY} Mocks}
\label{sc:Mocks}
In addition to the void catalogue, we also use the same set of voids found from 1000 mock galaxy catalogues as used in \citet{Woodfinden2022}. These mocks were generated using the \texttt{PATCHY} algorithm \citep{Kitaura2016}. Given a reference mock found from an HOD model applied to an N-body simulation, the \texttt{PATCHY} algorithm produces multiple mocks using augmented Lagrangian perturbation theory to model structure formation, and a bias model with deterministic and stochastic elements conditioned on reproducing the two- and three-point clustering functions. The end result is a mock galaxy catalogue that reproduces the monopole and quadrupole galaxy clustering signal of LOWZ and CMASS, as well as the angular footprint, masking effects, and redshift distribution of the voids detected in the real data. We use the cosmology values in the PATCHY mocks as our fiducial cosmology in this work for consistency. 

The goal of void catalogues found from these mocks is to construct sets of random lenses that will be used to measure and subtract out lingering shear systematics in the UNIONS shape catalogue, as described in \citet{Mandelbaum2005}. Further, as we will later choose to use an analytic model of our covariance, it was demonstrated in \citet{Singh17} that subtracting the signal around random lenses simplifies the covariance model. Lastly, we use the covariance across these random lens catalogues to validate some of the simplifications outlined in Sect.~\ref{sc:cov} for the analytic covariance.

\section{Methodology}
\label{sc:Methods}
\subsection{Weak Lensing Formalism}
\label{sc:weaklens}
In a weak lensing framework, the correlated shapes of background galaxies are directly related to the excess surface mass density (ESMD) $\Delta \Sigma$. We can recover the ESMD profile from source shapes by taking a weighted average over the shapes in each radial bin, $R$, around a void:
\begin{equation}
    \label{eq:DelSigdef}
    \Delta \Sigma \left(R\right)=\frac{\langle \gamma\sbr{t} \rangle \left(R\right)}{\sigcinv}\,,
\end{equation}
where $\sigcinv$ is the inverse of the critical density averaged over the source distribution
\begin{equation}
\label{eq:sigcrit}
    \sigcinv \left(z\sbrc{L}\right)=\frac{4\pi G\,\chi\left(z\sbrc{L}\right)\,\left(1+z\sbrc{L}\right)}{c^2}\int_{z\sbrc{L}}^\infty dz\sbrc{S}\; n\left(z\sbrc{S}\right)\frac{\chi\left(z\sbrc{L},z\sbrc{S}\right)}{\chi\left(z\sbrc{S}\right)} \,,
\end{equation}
where $n\left(z\sbrc{S}\right)$ is the redshift distribution of the sources, $\chi\left(z\sbrc{L}\right)$ is the comoving distance to the lens at redshift $z\sbrc{L}$, $\chi\left(z\sbrc{S}\right)$ is the distance to the source at redshift $z\sbrc{S}$, $\chi\left(z\sbrc{L},z\sbrc{S}\right)$ is the distance between the lens and source. All distances are measured in comoving units, chosen to facilitate easier modelling of the covariance, as described later. In the version of the UNIONS data used here, photometric redshifts are not available, so we integrate over the entire redshift distribution. 

In the case of many weak lensing applications, such as galaxy-galaxy lensing, a single lens does not yield a significant detection. This is also true of cosmic voids \citep{Amendola} and so, as suggested in \citet[][]{Krause2013}{}{} it is necessary to stack the voids by combining the source shapes behind all of the voids into a single set of radial bins and making a single measurement, effectively treating all of the sources as though they were lensed by a single, average void. 

However, when stacking voids, an additional issue arises that is not typically present in standard galaxy-galaxy or cluster lensing analyses. Voids have a wide range of radii, and so if we were to apply a universal set of radial bins specified in Mpc, for example, to each void, then the resulting average void density profile would be washed out, as the overdense walls of the smaller voids would be entirely contained within of the underdensities of the larger voids. Instead, we therefore scale the radial bins by the radius of each void. Doing so allows the distinct features of the profile to occur in roughly the same bins. The resulting measurement then becomes the weighted average of the shapes in these scaled bins:
\begin{equation}
    \Delta \Sigma \left(R/R\sbr{V}\right)=\frac{\sum\sbrc{L}\sum\sbrc{S} w\sbrc{LS}\gamma\sbrc{t,LS}\left(R/R\sbr{V}\right)\sigcinv^{-1}}{\sum\sbrc{L}\sum\sbrc{S} w\sbrc{LS}}\,,
\end{equation}
where the sum over S is the sum over all sources in a single radial bin across all lenses, and $w\sbrc{LS}$ is the weight assigned to each lens-source pair. The optimal weight for a shape-noise dominated regime is given by \citep{Sheldon_2004,Shirasaki2018} 
\begin{equation}
    \label{eq:lensing_weights}
   w\sbrc{LS}=\frac{\sigcinv^2}{2\sigma_\mathrm{\epsilon}^2+\sigma_{\epsilon_1}^2+\sigma_{\epsilon_2}^2}=\sigcinv^2w\sbrc{S} \,,
\end{equation}
using the shape weights $w\sbrc{S}$ as in Eq.~\eqref{eq:sourceweight}, with an additional factor of $\sigcinv^2$. This additional factor downweights lens-source pairs that are physically close to each other, where source galaxy shapes will not experience a strong lensing effect \citep{Shirasaki2018}. 

\subsection{Random Lensing Voids from \texttt{PATCHY} Mocks}
\label{sc:randcat}
It is common practice to subtract the lensing signal around randomly distributed points in the lens plane. This allows for the estimation and subsequent subtraction of lingering shear systematics that can spuriously correlate the shapes of the source galaxies \citep{Mandelbaum2005}. However, due to the fact that our radial bins scale with the physical size of the void, we must assign a radius and redshift to our random points. Because voids that touch the boundary of the BOSS survey are removed, we must also consider a selection bias towards smaller voids near the BOSS survey boundaries. This bias would not be present if we built a catalogue from a random distribution of points across the survey footprint and assigned each point a radius and redshift by sampling the corresponding data distributions. Furthermore, building a void catalogue by running \texttt{Revolver} on a random set of tracer points would be insufficient, as this catalogue would fail to reproduce the void size distribution in the data.

Fortunately, void catalogues produced by running the same \texttt{Revolver} void finder on the \texttt{PATCHY} mocks will reproduce these edge effects, so they can be used to generate a set of realistic randomly distributed lenses. This produces a set of 1000 mock void catalogues. However, if we were to sample random voids entirely within a single mock catalogue, since each mock attempts to reproduce the clustering signal in the galaxy surveys, it would necessarily approximate the clustering of the voids present in the data. This clustering signal needs to be removed so that the mocks can mirror a uniform random distribution in position while maintaining the selection biases at the boundaries. 

To remove this clustering signal and build our random lens catalogues, we randomly pick out individual voids from across all 1000 mocks until we construct a catalogue that matches the number of voids in each data set. We then do this 30 times to construct 30 separate random catalogues per data set. Then, we measure the stacked $\Delta\Sigma$ signal from each of the catalogues. This results in 30 independent measurements of systematic shape correlations in each radial bin. Then we take an average across all of the random measurements in each bin, as well as compute the $14\times14$ covariance matrix across the 14 radial bins using the 30 measurements. This covariance, which represents an empirical estimate of the covariance expected in the data, will be used to validate our covariance model in Sect. \ref{sc:cov}.

\begin{figure}
    \centering
    \includegraphics[width=\columnwidth]{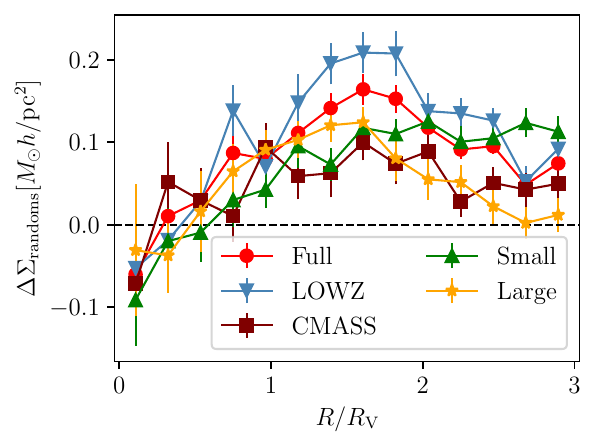}
    \caption{Results from measuring the $\Delta\Sigma$ profile around 30 sets of random points from mocks. Each data point is an average of the 30 random catalogues for each radial bin, and the error bars are the square root of the diagonal of the covariance matrix found from the randoms divided by the 30 catalogues that went into the measurement.}
    \label{fig:LFcomb_randoms}
\end{figure}

The measured $\Delta \Sigma$ profiles around randomly distributed lenses for all catalogues can be seen in Fig.~\ref{fig:LFcomb_randoms}. The error bars shown in this figure represent the 1$\sigma$ scatter in each bin across all 30 catalogues, reduced by the factor $1/\sqrt{30}$ to represent the error in the mean. Since the lens centres for these data sets are distributed randomly, we expect that there should not be any coherent lensing signal present in these results. However, we do find a consistent trend across all but the smallest scales where the measured values are larger than zero, pointing towards low-level systematics that are nevertheless large enough (compared to the weak void measurement, roughly 10 to 20 per cent of the signal) that they will need to be subtracted off. We do note however, that these systematics are of the order $4\times10^{-5}$ when converted to tangential shear, and would represent a small contribution to more standard applications of lensing. 

To test the effects of this methodology over a simple random sample, we measured the signal around a uniform random sample of voids, with void radii sampled from the distribution of void radii in the actual LOWZ voids. Additionally, to see if this signal is due to the bin scaling required by voids, we measured the signal around a uniform random sample of galaxies, chosen in the same way as the uniform random voids, just without specifying a radius. Both uniform samples were chosen with properties of the LOWZ catalogue as it has the largest amplitude random ``signal''. We found that there still exists a non-zero signal, but at a roughly constant value of around $0.05\,M_\odot h^{-1} \mathrm{pc}^{-2}$, consistent between random voids and random galaxies, indicating the bin scaling does not have a strong effect on the random ``signal''. Comparing this low and constant value from a uniform random sample to the non-constant and larger amplitude signal seen in LOWZ in Fig.~\ref{fig:LFcomb_randoms}, which comes from mock voids, we conclude that the selection effects at the boundary captured by the mocks has an effect on the measured systematics.

To obtain our final measurements, we subtract the randoms (shown in Fig.~\ref{fig:LFcomb_randoms}) from the data. We further add the covariance matrix obtained from the randomly distributed lenses, divided by 30 (representing the covariance in the mean), to the analytic covariance discussed in Sect.~\ref{sc:cov} to propagate these errors to our final measurement. 
\subsection{Analytic Covariance Modelling}
\label{sc:cov}
Obtaining an accurate estimate for the void lensing covariance is essential because the signal itself is weak, as demonstrated by other analyses \citep[e.g.\ ][]{Melchior2014,Fang2019}, and also due to the nature that the scales considered are large. The total number of lenses is also relatively low, indicating that large-scale structure (LSS) effects are expected to have a stronger presence in the error budget of this form of lensing analysis. This differs from the more familiar case of galaxy-galaxy lensing, where the small scales probed by galaxy lenses, coupled with the large number statistics of lens galaxies, usually means that covariance contributions from LSS are subdominant to the purely diagonal shape noise component. On the other hand, the large scales of cosmic voids mean that the covariance is insensitive to nonlinear scales, simplifying the modelling. To this end, we opt to analytically model the covariance for void lensing. This is the first time an analytic covariance matrix has been used for a void lensing analysis with data. We start with the analytic covariance formula from \cite{Krause2013}
\begin{multline}
    \label{eq:krause_cov}
    \textrm{Cov}\left[\gamma\sbr{t}\left(\theta\sbr{1}\right),\gamma\sbr{t}\left(\theta\sbr{2}\right)\right]=\frac{1}{4\pi f\sbr{sky}}\int \frac{\ell d\ell}{2\pi}\widehat{J_2\left(\ell\theta\sbr{1}\right)}\widehat{J_2\left(\ell\theta\sbr{2}\right)}\\
    \left\{\left[C\sbr{VV}\left(\ell\right)+\frac{1}{n\sbr{V}}\right]\left[\ckk\left(\ell\right)+\frac{\sigma_{\epsilon}^2}{n\sbr{g}}\right]+\left[\cvk\left(\ell\right)\right]^2\right\} \,,
\end{multline}
where $f\sbr{sky}$ is the sky fraction of the survey, $\widehat{J_2}$ is the 2nd order Bessel function of the first kind averaged over each radial bin, $n\sbr{V}$ and $n\sbr{g}$ are the angular number density of voids and source galaxies, respectively. $\theta_1$ and $\theta_2$ are defined to be the angular extent of the radial bins, which we calculate by multiplying the corresponding $R_1/R\sbr{V}$ and $R_2/R\sbr{V}$ values by the angular radius of the void $\theta_V$ to convert to angular units. The terms $C\sbr{VV}$, $\ckk$, and $\cvk$ are the angular power spectra representing the clustering of voids, source galaxy shapes, and their cross-correlation, respectively. This covariance formula is similar to the Gaussian component of covariances used in standard lensing analyses, only using voids as a tracer population \citep[e.g.\ ][]{Jeong09, Singh17, Krause17}{}{}. Per \citet{Singh17}, this form of the covariance matrix only applies once the tangential shear around random points is subtracted from the overall signal. Otherwise, additional terms would need to be added in to account for the fact that we would be correlating a void field (which does not have a mean of zero) with a mean-zero shear field. Subtracting the random signal turns the correlation into a correlation between mean-zero quantities.

We demonstrate in Appendix~\ref{sc:appendix_cov} that terms involving $\cvk$ and $C\sbr{VV}$ are subdominant to the $\ckk/n\sbr{V}$ term across all scales, and can therefore be dropped. To measure the variance due to shape noise (which arises from the product of $1/n\sbr{V}$ and $\sigma_\epsilon^2/n\sbr{g}$), we compute the error in the weighted mean:
\begin{equation}
    \sigma^2_{\mathrm{SN,}\Delta\Sigma}= \frac{\sum\sbrc{L}\sum\sbrc{S} w\sbrc{LS}^2\sigcinv^{-2}\sigma_\epsilon^2}{\left(\sum\sbrc{L}\sum\sbrc{S}w\sbrc{LS}\right)^2}\,,
\end{equation}
where the $\Delta\Sigma$ subscript indicates that we include a factor $\sigcinv^{-2}$ to convert the shear to $\Delta\Sigma^2$ units, and $\sigma_\epsilon=0.34$ as measured in the UNIONS survey. These assumptions yield the much simpler formula:
\begin{multline}
    \textrm{Cov}\left[\gamma\sbr{t}\left(\theta\sbr{1}\right),\gamma\sbr{t}\left(\theta\sbr{2}\right)\right]=\frac{1}{4\pi f\sbr{sky}}\int \frac{\ell d\ell}{2\pi} \widehat{J_2\left(\ell\theta\sbr{1}\right)}\widehat{J_2\left(\ell\theta\sbr{2}\right)}\left[\frac{\ckk\left(\ell\right)}{n\sbr{V}}\right]\\+\delta^\mathrm{K}_{\theta_1,\theta_2}\sigma_{\mathrm{SN}}^2\,,
    \label{eq:cov_simplify}
\end{multline}
where $\delta^{K}$ is the Kronecker delta function, to represent that the measured shape noise only adds a diagonal contribution to the covariance. Using the Limber approximation \citep{Limber53,Loverde08}
\begin{equation}
\label{eq:ckk}
    \ckk\left(\ell\right)=\int d\chi\left[\frac{\bar{\rho}}{\chi}\int_\chi^\infty dz\sbrc{S}\,\frac{c}{H\left(z\sbrc{S}\right)}\frac{p\left(z\sbrc{S}\right)}{\Sigma\sbr{c}\left(z\sbrc{S},z\right)}\right]^2P\sbr{mm}\left(k=\frac{\ell+1/2}{\chi}\right)\,,
\end{equation}
with $1/\Sigma\sbr{c}$ is defined as in Eq.~\eqref{eq:sigcrit} without the integral over the source distribution. As mentioned earlier, because the scales we probe with voids are quite large, we only need to use a linear matter power spectrum model. 

The remaining $\ckk/n\sbr{V}$ term in the covariance formula physically represents covariance between radial bins due to lensing from intervening large-scale structure along the line of sight. It is then intuitive why this is the dominant term in the LSS component of the covariance: the combination of the sparseness of the LRG tracers resulting in very large voids, coupled with the low redshifts of the BOSS sample, creates bins that can extend upwards of eight degrees (using the average void radius) in radius on the sky. The chances of large-scale structure existing within these bins along the line of sight in a single pointing are then reasonably high. Furthermore, the relatively low number density of foreground voids results in higher shot noise compared to analogous applications of this covariance formula, like galaxy-galaxy lensing.

However, we note that Eq.~\eqref{eq:cov_simplify} is the covariance of the lensing shear $\gamma_t$, and not the covariance of the excess surface density $\Delta \Sigma$. Following the examples of \citet{Shirasaki2018} and \citet{Wu2019}, who modified this covariance matrix for cluster lensing, we need to include an additional factor of $\Sigma\sbr{c}^2\left(z\sbrc{S},z\sbrc{L}\right)$ into Eq.~\eqref{eq:ckk}. Since there is a remaining integral over source redshift present in Eq.~\eqref{eq:ckk}, normally, this extra factor would need to be included in this integrand. However, this integral is not well-behaved for sources in front of or at the same redshift as the lenses, $z\sbrc{L}$. As there is no method to identify which sources these are without some estimate of individual redshifts,  we are not computing the same measure of the excess surface density as \citet{Wu2019} or \citet{Shirasaki2018}, as their lensing estimator has $\Sigma\sbr{c}$ is an explicit function of the source redshift, while ours is expressed as an average over the source distribution, as seen in Eqs.~\eqref{eq:DelSigdef} and \eqref{eq:sigcrit}. Since $\sigcinv$ no longer depends on the source redshift, it can factor out of both integrals in Eq.~\eqref{eq:ckk}. It can also factor out of the integral of Eq.~\eqref{eq:cov_simplify}, leaving
\begin{multline}
\label{eq:cov_dsig}
    \textrm{Cov}\left[\Delta \Sigma\left(\theta\sbr{1},z\sbrc{L}\right) \Delta \Sigma\left(\theta\sbr{2},z\sbrc{L}\right)\right]= \frac{1}{4 \pi f\sbr{sky}\langle \Sigma^{-1}\sbr{c}\rangle^2\left(z\sbrc{L}\right)} \\ \int \frac{\ell d\ell}{2\pi}\widehat{J\sbr{2}\left(\ell\theta\sbr{1}\right)}\widehat{J\sbr{2}\left(\ell\theta\sbr{2}\right)}\left[\frac{\ckk\left(\ell\right)}{n\sbr{V}}\right]+\delta^\mathrm{K}_{\theta_1,\theta_2}\sigma_{\mathrm{SN,}\Delta\Sigma}^2\,.
\end{multline}

Finally, because our measurement uses radial bins that depend on each individual void's radius, there is an ambiguity in terms of what angular scales to consider for the Bessel window functions, as each void effectively probes different scales of the LSS covariance term. This is in contrast to traditional weak lensing setups, where each lens uses the same set of bins, so $\theta_1$ and $\theta_2$ are consistent between all lenses. There are two main ways to account for this. One method would be to bin our void catalogue by void radius, reducing this ambiguity by reducing the spread in void size. However, the goal of this work is to maximize the level of significance of our detection, so we want to avoid splitting the void lensing information too finely. The other method which we chose to perform is to compute the covariance for each void individually. We first note that in the reduced form of the covariance equation in Eq.~\eqref{eq:cov_dsig}, we can bring $(4\pi f\sbr{sky})^{-1}$, which is the inverse of the survey area, into the integral and multiply it by the inverse area density of voids, $n\sbr{V}^{-1}$. This results in the inverse total number of voids, $N\sbr{V}^{-1}$. For a single void, we set this to one, and in this way compute the covariance per void. We can then sum the covariances and weight them in the exact same way we did for the shape noise
\begin{equation}
\label{eq:cov_stack}
    \textrm{Cov}\left[\Delta\Sigma\left(\theta_1,z\sbrc{L}\right)\Delta\Sigma\left(\theta_2,z\sbrc{L}\right)\right]=\frac{\sum_Lw\sbrc{L}^2f\left(z\sbrc{L},\theta_1,\theta_2\right)}{\left(\sum_Lw\sbrc{L}\right)^2}+\delta^\mathrm{K}_{\theta_1,\theta_2}\sigma^2_{\mathrm{SN,}\Delta\Sigma}\,,
\end{equation}
where
\begin{equation}
\label{eq:cov_individual}
    f\left(z\sbrc{L},\theta_1,\theta_2\right)=\frac{1}{\langle \Sigma\sbr{c}^{-1}\rangle^2\left(z\sbrc{L}\right)}\int \frac{\ell d\ell}{2\pi}\widehat{J\sbr{2}\left(\ell\theta\sbr{1}\right)}\widehat{J\sbr{2}\left(\ell\theta\sbr{2}\right)}\left[\frac{\ckk}{1}\right]\,,
\end{equation}
and $w\sbrc{L}$ is the lens-source weights defined in Eq.~\eqref{eq:lensing_weights} summed over all sources belonging to each lens.
\begin{figure}
    \centering
    \includegraphics[width=\columnwidth]{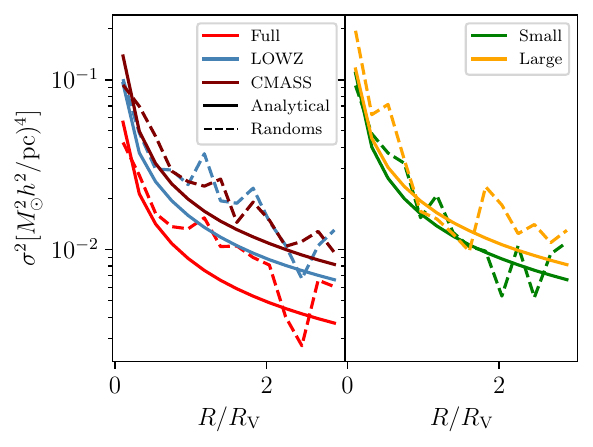}
    \caption{Comparison between the diagonals of the analytic covariance calculation and the empirical estimate of the covariance around mock-random lenses for the Full void catalogue and the four additional subsets used in this analysis.}
    \label{fig:covcompare}
\end{figure}

In principle, one can use the covariance of the signal found using the random catalogues as an empirical estimate of the covariance matrix. This matrix would only be sensitive to the shape noise and $\ckk$ terms due to the use of random unclustered lenses. However, we find that the covariance from randoms in the off-diagonal elements is too noisy to be used in lieu of this model for the analysis. This is largely due to correlating random voids with sources from data. Given that voids are very large objects, we quickly begin to oversample the footprint of the survey overlap with increasing numbers of randoms. We therefore find that increasing the number of random catalogues from 30 to 50 does not significantly dampen the noise. 

However, the signal in the diagonal elements is larger than the typical fluctuations in the off-diagonal elements ($<0.01$), so the diagonal components of the random covariance matrix can be used as a comparison test to help validate our model, as our model is similarly insensitive to the $\cvk$ and $C\sbr{VV}$ terms. Since our model does not include any terms that only contribute to the off-diagonal elements, testing the diagonal alone would be sufficient. Fig.~\ref{fig:covcompare} shows the comparison of the diagonals of the matrices from Eq.~\eqref{eq:cov_stack} and from the randoms. We can see good agreement across samples and across bins, with the empirical estimate only having a large disagreement with the model at the smallest bin, which is likely due to scatter from a large shot noise due to the smallest bin having the fewest galaxies. This result, coupled with the result in Appendix~\ref{sc:appendix_cov}, then validates the approximations that go into our covariance model.

\subsection{Void Density Profile Models}
\label{sc:Profile_Header}
\subsubsection{Analytic Void Profile}
\label{sc:hsw}
The primary result from this measurement will be an estimate of the excess surface mass density, defined in terms of the surface mass density $\Sigma$ as
\begin{equation}
    \Delta \Sigma\left(R/R\sbr{V}\right)=\overline{\Sigma\left({<}R/R\sbr{V}\right)}-\Sigma\left(R/R\sbr{V}\right) \,,
    \label{eq:ESMD}
\end{equation}
where $\overline{\Sigma\left({<}R/R\sbr{V}\right)}$ is the average of the surface mass density within a circle of radius $R/R\sbr{V}$.

To obtain $\Sigma$, the three-dimensional mass density profile is integrated along the line of sight. For spherically symmetric profiles, such as the case of a void profile averaged over many voids, this can be simplified to an Abel transform
\begin{equation}
    \Sigma \left(R\right) = 2\int_R^\infty \frac{\rho\left(r\right)r}{\sqrt{R^2-r^2}}dr \,.
\end{equation}
For a particular model $\rho\left(r\right)$, we can calculate its corresponding $\Delta \Sigma$ profile, and then fit the model to the data. In the literature, there is one popular choice of empirical void profile presented in \citet[hereafter HSW]{Hamaus2014} \begin{equation}   
\label{eq:B+15}
    \frac{\rho\sbr{V}\left(r\right)}{\overline{\rho}}-1=\delta\left(r\right)=\delta\sbr{c}\frac{1-\left(r/r_\mathrm{s_1}\right)^\alpha}{1+\left(r/r_\mathrm{s_2}\right)^\beta} \,.
\end{equation}
This model as originally proposed has four free parameters: the central underdensity $\delta_c$, the zero-crossing scale $r_\mathrm{s_1}$, and the power-law parameters $\alpha$ and $\beta$, with the $r\sbr{s_2}$ parameter fixed to the void radius $R\sbr{V}$. We follow the modification proposed by \citet[hereafter B+15]{Barreira2015} and free the $r\sbr{s_2}$ parameter. This parameter then controls the scale at which the profile asymptotically approaches zero. The HSW model was found to fit the void profile of simulated voids found by the \texttt{VIDE} void finder, and has been found to fit void profiles from both data and simulations with a variety of void finding setups \citep{Chan2014,Barreira2015,Sanchez16,Pollina17,Falck2018,Baker2018,Pollina19,Fang2019,Shim2021,Tavasoli2021,Williams2025,Davies2025}.

We make note that the voids we consider in this analysis are circumcentre voids, which differ from barycentre voids found by \texttt{VIDE}. This is notable because circumcentres will by definition find the centre of the largest empty sphere, which can be located anywhere inside of the void. Since the sphere is empty, it is expected that the corresponding matter profile will have a steeper slope inside this sphere around the centre. Barycentre voids on the other hand will not have this feature. This methodological difference may require a modification to the HSW model. There have been some studies that have modified the HSW profile to account for differences in void finding methodology and modified gravity \citep{Barreira2015,Chantavat2017,Falck2018,Baker2018,Perico2019,Williams2025,Davies2025}. Additionally, for \texttt{Revolver} specifically, \citet{Nadathur2015} qualitatively find that the five-parameter modification by B+15 produces better visual fits to simulated \texttt{Revolver} void profiles.

We investigate the difference in fits between the four- and five-parameter HSW model applied to our voids further in Appendix~\ref{sc:HSW}. There, we find that these two models are relatively interchangeable (with at most a 1.9$\sigma$ significance improvement of the fit quality for the Full catalogue down to $<1\sigma$ in the CMASS and Small catalogues). However, we find that the original four-parameter HSW model produces best-fit values of $\alpha<1$ and $\delta\sbr{c}=-1$ across all catalogues. This produces void profiles with concave behaviour in the innermost regions. This is a property that has not been directly observed in the literature, as void profiles are typically flat to convex near their centre. This is likely due to a more fundamental limitation in the void-galaxy lensing methodology generally. Since the lensing profile relies on the number of source galaxies in a given bin, the innermost bins will have the fewest galaxies, therefore the innermost regions will not be well constrained. Since we are directly measuring $\Delta\Sigma$, large error bars in the smallest bins correspond to a large spread in possible inner slopes. We find that the five-parameter HSW model following the modification in B+15 produces best-fit void profiles that are somewhat more agreeable with expectations with $\alpha\geq1$ and $\delta\sbr{c}>-1$ at the cost of a more complex posterior region as investigated in Appendix~\ref{sc:HSW}. We therefore proceed with presenting results from the five-parameter HSW model, but note that the unmodified four-parameter HSW fits are still acceptable.

\subsubsection{Void-Galaxy Cross Correlation Measurement}
\label{sc:void-gal measure}
Assuming a scale-independent linear bias relation, we can write the void-matter cross-correlation as a function of the void-galaxy cross-correlation divided by a bias factor $b\sbr{Vg}$. How this $b\sbr{Vg}$ factor relates to the global linear $b\sbr{g}$ can then be used to better understand the dependence of galaxy bias on environment by examining its relationship to the least dense parts of the Universe. To this end, we use the results and methodology of \citet{Woodfinden2022}, with slight modifications, to measure the void-galaxy cross-correlation with the intent to construct a model with a free parameter $b\sbr{Vg}$ that will be constrained by fitting this model to our void lensing measurement. We refer readers to \citet{Woodfinden2022} for additional details and systematics tests for the void-galaxy cross-correlation. 

The measurement of the void-galaxy cross correlation is performed entirely within the North Galactic Cap (NGC) component of the LOWZ and CMASS surveys, in contrast with the void lensing measurement, where we consider the overlap of UNIONS and either LOWZ or CMASS. The galaxy data set used is the same LRG data set used in the void-finding procedure. We note that this means this galaxy data set is also in reconstructed space with approximate RSDs removed, and is therefore dependent on the same fiducial choice of $\beta=f/b=0.37$ as the location of the void centres. The cross correlation is measured using the \cite{Landy1993} estimator.

To account for the same selection bias as considered in the random catalogues used in the void lensing measurement, the void randoms for the void-galaxy cross-correlation are constructed from a subset of 250 of the 1000 \texttt{PATCHY} mocks. The void randoms are then randomly chosen until a catalogue exists with 50 times more randoms than data points for the purposes of the void-galaxy cross correlation measurement. Furthermore, we opt to use the void-galaxy cross correlation measurement in reconstructed space rather than the redshift space correlation that was the primary result of \citet{Woodfinden2022}. Correlating with the galaxy field in reconstructed space will in principle produce a more spherically-symmetric profile that will better represent the isotropy expected from the average void lensing profile.

\begin{figure}
    \centering
    \includegraphics[width=\columnwidth]{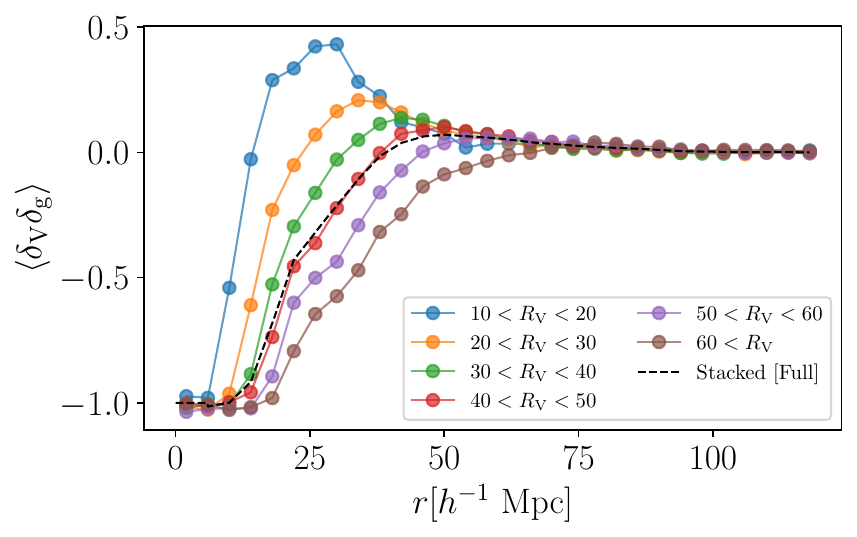}
    \caption{BOSS void-galaxy cross-correlation measurements binned by void radius. The black dashed line represents the weighted average across all binned profiles for the Full catalogue. This average is then used as the void-galaxy model for the void lensing measurements to estimate galaxy bias. These measurements use the entirety of the BOSS void catalogue in the North Galactic Cap (NGC) as opposed to the region that overlaps UNIONS considered in the void lensing analysis.}
    \label{fig:vg_cross_corr}
\end{figure}

We note that in the case of both void lensing and void-galaxy measurements, we are computing a weighted average across all voids in the sample. We therefore want to ensure that we choose a similar weighting scheme between both measurements. To do this, we slightly modify the setup of \citet{Woodfinden2022}. We do not perform a minimum radius cut on the void catalogue, and instead bin voids by void radius, in steps of 10 $h^{-1}$Mpc. This radius cut in  \citet{Woodfinden2022} was done for the purposes of analysing the RSD signal, and as such is not needed for our analysis. We then perform the void-galaxy measurement on each of these radius bins, the results of which are plotted in Fig.~\ref{fig:vg_cross_corr}.
We then build an effective void size distribution from the lensing catalogues (which are subsets of the total NGC void sample), weighted by the lensing weights. We choose the bins of this distribution to match the $10\,h^{-1}$Mpc bins used in the void-galaxy measurement. We then take the average of the void-galaxy profiles shown in Fig.~\ref{fig:vg_cross_corr}, using this void size distribution as weights. This procedure will approximate the weighting scheme used in the lensing analysis, since not all of the NGC void catalogue are in the void lensing catalogues. The resulting void-galaxy profiles binned by void radii are shown in Fig.~\ref{fig:vg_cross_corr}, where this average profile used for the Full catalogue is shown in the dashed line. It should be noted that these void-galaxy cross-correlation measurements were performed in unscaled bins in the same manner as \citet{Woodfinden2022} and, as such, will not span the entire range of radial bins used in the void lensing measurement.

For the void-galaxy measurements shown in Fig.~\ref{fig:vg_cross_corr}, we note a few caveats that pertain especially to the innermost regions. In general, the density is difficult to measure there due to several factors: the effects of shot noise are high precisely because the density is low, the density will depend on the method used to estimate it (e.g.\ does one simply count zero galaxies as $\delta = -1$, or does one use a different density estimator such as the inverse of the volume of Voronoi cells), as well as the definition of the void centre. Since here we are considering circumcentre voids, where the centre is maximally located away from all galaxies within the void, it must be the case (i.e.\ it is by construction) that the central galaxy underdensity is $-1$. Furthermore, there exists a kink in the profiles at around 20 to 25 $h^{-1}\,\mathrm{Mpc}$ that exists because of a sharp density transition that occurs between the galaxies used for determining the circumcentre and the location of the centre \citep{Nadathur2019b}. This constructed lack of pair counts near the centre of circumcentre voids may not be sensitive to the linear bias between the galaxy density and the matter density when using a pair count estimator. In contrast, \citet{Pollina17} considers barycentre voids, which have a smoother profile near the centre. This difference can partly explain the amplitude of the galaxy bias residual found in \citet{Nadathur2019b}, however they additionally find that this scale-dependent bias residual also exists for barycentre voids. On the other hand, \citet{Curtis2025}, using the \texttt{TNG300} simulation find galaxy bias terms from circumcentre voids that are fit well by the linear bias model when compared to dark matter profiles of voids in the range of 0.1 to 1.2 $R\sbr{V}$. The exception occurs at high redshift and high minimum stellar mass cuts, where significant deviations are likely due to increased tracer sparsity, highlighting the above caveats. While the highest minimum stellar mass cut considered in \citet{Curtis2025} is an order of magnitude lower than the expected stellar mass of CMASS galaxies, the authors do find good agreement with the linear bias model at the redshift range of LOWZ and CMASS using circumcentre voids. Therefore, we still expect fitting these void-galaxy cross-correlations to the void lensing profile across all bins to be sensitive to the linear bias.

In order to convert this measurement into a model that can fit our void lensing measurement, we will assume the linear bias relation
\begin{equation}
\label{eq:bias_relation}
    \langle \delta\sbr{V} \delta\sbr{g}\rangle\left(r\right)=\xi\sbr{Vg}^\mathrm{3D}\left(r\right)=b\sbr{Vg}\langle \delta\sbr{V} \delta\sbr{m}\rangle\left(r\right) = b\sbr{Vg} \, \xi\sbr{Vm}^\mathrm{3D}\left(r\right)  \,,
\end{equation}
and so 
\begin{equation}
 \xi\sbr{Vm}^\mathrm{3D}\left(r\right) = \frac{\xi\sbr{Vg}^\mathrm{3D}\left(r\right)}{ b\sbr{Vg}} \,.
\end{equation}
We can thereby treat $1/b\sbr{Vg}$ as a free parameter to scale the amplitude of the void-galaxy cross-correlation measurement to fit our measured void lensing signal. We differentiate this value from $b\sbr{g}$, which is the large-scale linear galaxy bias, as the values are expected to differ (at least for smaller voids). However, like the analytic models, the void-galaxy measurements are done in 3D space, while void lensing is done in units of excess surface mass density. Starting with Eq.~\eqref{eq:bias_relation}, we can take the projection of both sides to arrive at
\begin{equation}
    \Sigma\left(R\right)=\frac{2\overline{\rho}}{b\sbr{Vg}}\int_R^\infty \frac{\xi_\mathrm{Vg}\left(r\right) \, r}{\sqrt{R^2-r^2}}dr\,,
\end{equation}
where $\overline{\rho}$ denotes the average density of the Universe, which converts the dimensionless void-galaxy cross correlation into the same units as the ESMD measurement. Then, plugging this equation into Eq.~\eqref{eq:ESMD}, we can relate $\xi\sbr{Vg}$ to $\Delta\Sigma$ from our void lensing measurement with a free parameter of $b\sbr{Vg}$.

\subsubsection{Model Fitting Methodology}
\label{sc:model_fit}
For both the HSW and void-galaxy models, we perform a least-squares fit by minimizing $\chi^2$
\begin{equation}
\label{eq:chisq}
    \chi^2=\sum_{i,j}\left(\Delta\Sigma_{i\mathrm{,data}}-\Delta\Sigma_{i\mathrm{,model}}\right)C^{-1}_{ij}\left(\Delta\Sigma_{j,\mathrm{data}}-\Delta\Sigma_{j\mathrm{,model}}\right)\,.
\end{equation}
The resulting parameter fits for the HSW model are then used to determine the significance of the measurement. This is done by comparing the $\chi^2$ of the best-fit model to the $\chi_0^2$ of a model set to zero in every bin, which represents a null detection
\begin{equation}
    \Delta\chi^2=\chi^2_0-\chi^2\,.
\end{equation}
Using Wilks' Theorem \citep{Wilks38}, the distribution of this $\Delta\chi^2$ statistic is itself a $\chi^2$ distribution, with degrees of freedom equal to the difference in degrees of freedom between the two nested models, which in our case for the modified HSW model is five. The area under the tail of this distribution then carries information about the significance of the detection, and serves as one of the primary results of this work. For the void-galaxy model fits, the resulting parameter fit is a measurement and uncertainty of $b\sbr{Vg}$, the galaxy bias parameter in underdense environments, another primary result of this work.

We additionally perform a MCMC analysis to the HSW model to explore the posterior around the best-fitting parameters. This was performed using the \texttt{emcee} \citep{emcee} library minimizing the log likelihood, converted from Eq.~\eqref{eq:chisq} by a factor of $-0.5$. For both the MCMC and least-squares fits of the HSW model, we apply a set of priors on the five different parameters. For $\delta_\mathrm{c}$, $r_\mathrm{s_1}$, and $r_\mathrm{s_2}$, these are uniform flat priors. For $\alpha$, we enforce $\alpha\geq1$, as the model goes as $\delta\sbr{c}\left(1-r^\alpha\right)$ on small scales. As mentioned previously, if $\alpha<1$, then the matter density profile near $r=0$ is concave, a behaviour not seen in previous works, which have all shown void profiles with convex behaviour, i.e.\ whose initial central slope increases with radius. 

Additionally, looking at the large-scale behaviour of the HSW model, the density scales as $r^{\mathrm{\alpha-\beta}}$, and so this combination controls the rate at which the 3D matter profile drops to 0. In performing initial fits to the Full, LOWZ, and Large void catalogues, we initially tested a prior of $\alpha\leq\beta$, as this prevents models with infinite density at large scales. However, it was found that these catalogues in particular prefer fits with $\alpha\approx\beta$, which results in nonphysical 3D matter profiles with constant density out to the largest scales. This prompted the use of a stronger prior where we rely on the fact that small voids tend to have overcompensated profiles, where the matter in the overdense wall is more than is required to ``fill in'' the underdensity, while large voids result in compensated or undercompensated profiles. This implies that smaller voids will have a stronger overdensity signal at larger $r/R\sbr{V}$ bins than bigger voids. Therefore, using the best-fit model of the Small void catalogue, we construct a prior where all model parameters that result in a density profile with a value larger than the Small catalogue at a sufficiently large radius (the value at 5 $r/R\sbr{V}$ was adopted) are considered to be outside of the prior bounds and therefore rejected, effectively setting the scale where all profiles should be asymptotically close to 0. To test how reasonable this prior is, we can fit the void-galaxy measurements with the HSW profile to quantify their large-scale trends. Since we expect the void-galaxy and void lensing profiles to be similar, this gives us an order of magnitude test for the expected values of $\alpha-\beta$. We find that this choice of prior results in void lensing profiles with slightly lower values of $\alpha-\beta$ than values found in the void-galaxy profiles. This indicates that the void lensing profiles return to zero at large scales slightly faster, but at a relatively similar rate as the void-galaxy profiles, supporting the use of this prior. 
\section{Results}
\label{sc:results}
\renewcommand{\arraystretch}{1.5}
\begin{table*}
    \centering
    \caption{Fit statistics and significance measurements for the 5 void catalogues used in this work. The models are the five-parameter HSW analytical formula, also used to determine the significance (Sig.) of the measurement, and the void-galaxy (VG) cross-correlation measurement, used to measure the galaxy bias in underdense environments. Quantities with error bars represent the mean of the projected MCMC posteriors with 1$\sigma$ error bars. Quantities in parentheses represent the best-fitting model in the multidimensional posterior. Skewed posterior distributions are presented with upper or lower bounds, and represent the 95 per cent or 5 per cent upper or lower bound, respectively.}
    \begin{tabular}{|c|c|c|l|l|l|l|l|c|c|c|}
    \hline
     & \multicolumn{7}{c|}{Five-Parameter HSW Model} & \multicolumn{2}{c|}{VG Model} \\
     \hline
     Sample & $\chi^2$/dof & Sig. & \multicolumn{1}{c|}{$\delta\sbr{c}$} & \multicolumn{1}{c|}{$r_\mathrm{s_1}/R\sbr{V}$} & \multicolumn{1}{c|}{$r_\mathrm{s_2}/R\sbr{V}$} & \multicolumn{1}{c|}{$\alpha$} & \multicolumn{1}{c|}{$\beta$} & $\chi^2\sbr{Vg}$/dof  & $b\sbr{Vg}$ \\
     \hline
        Full & 15.6/9 & 6.2$\sigma$ & $-0.30^{+0.09}_{-0.04}\,,\left(-0.48\right)$ & $1.21_{-0.54}^{+0.62}\,,\left(0.81\right)$ & $0.63_{-0.08}^{+0.05}\,,\left(0.74\right)$ & ${<}14.8\,,\left(1.2\right)$ & ${>}7.3\,,\left(6.0\right)$ & 10.2/10 & $2.47\pm 0.36$\\
        LOWZ & 20.2/9 & 4.4$\sigma$ & $-0.28^{+0.10}_{-0.05},\, \left(-0.33\right)$ & $1.19^{+0.80}_{-0.53}\,,\left(0.77\right)$ & $0.65^{+0.07}_{-0.10}\,,\left(0.91\right)$ & ${<}15.1\,,\left(2.4\right)$ & ${>}8.03\,, \left(20.0\right)$ & 17.4/10 & $2.49\pm 0.49$\\
        CMASS & 12.3/9 & 3.4$\sigma$ & $-0.39^{+0.19}_{-0.10}\,,\left(-0.39\right)$ & ${>}0.73\,,\left(0.90\right)$ & $0.52^{+0.08}_{-0.16}\,,\left(0.85\right)$ & ${<}7.4\,,\left(1.1\right)$ & $7.9^{+2.1}_{-0.8}\,,\left(6.2\right)$ & 9.5/10 & $2.48\pm 0.55$\\
        Small &  5.7/9 & 4.5$\sigma$ & $-0.33^{+0.16}_{-0.09}\,,\left(-0.39\right)$ & $0.75^{+0.04}_{-0.06}\,,\left(0.71\right)$ & $0.93_{-0.13}^{+0.13}\,,\left(0.87\right)$ & ${<}8.8\,,\left(2.0\right)$ & $12.1^{+2.0}_{-4.8}\,,\left(7.3\right)$ & 15.1/13 & $2.82\pm 0.60$\\
        Large & 11.8/9 & 4.4$\sigma$ & $-0.24^{+0.08}_{-0.05}\,,\left(-0.30\right)$ & ${>}0.81\,,\left(1.37\right)$ & $0.67^{+0.06}_{-0.11}\,,\left(0.66\right)$ & ${<}15.0\,,\left(1.2\right)$ & ${>}6.9\,,\left(5.3\right)$ & 14.3/8 & $2.77\pm 0.57$\\
        \hline
    \end{tabular}
    \label{tab:fit_stats}
\end{table*}
\renewcommand{\arraystretch}{1}

\subsection{Void profiles}
\label{sc:Results_profiles}
\begin{figure*}
    \centering
    \includegraphics[height=9 in]{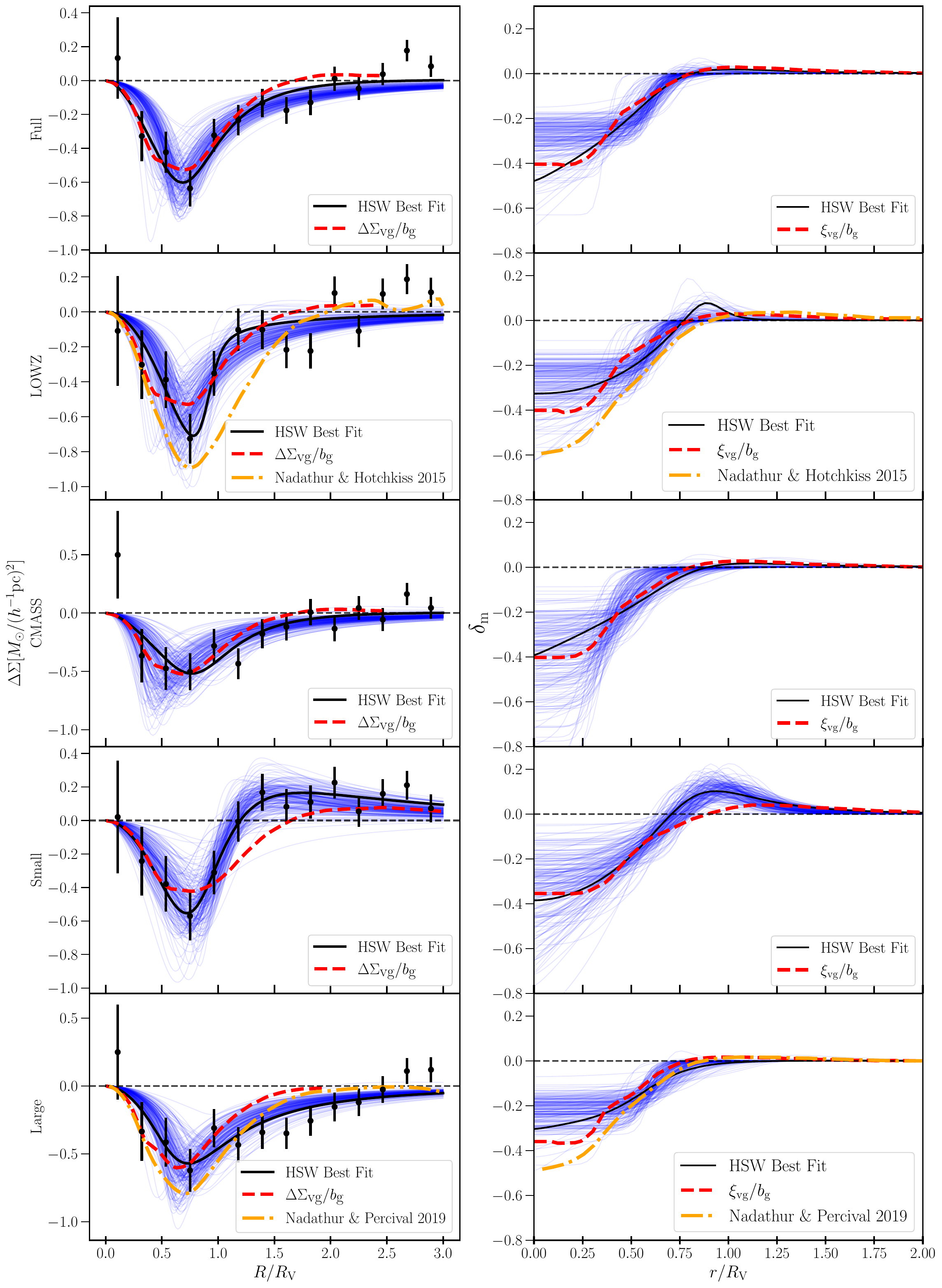}
    \caption{Void lensing measurements in $\Delta \Sigma$ units on the left, and the corresponding 3D density profiles on the right. These measurements are fit with two models: the analytic five-parameter HSW model in the black solid line, and the best fit bias scaling of the void-galaxy cross-correlation measurement. The thin blue lines show 200 draws from the HSW model MCMC chains. Orange dot-dashed lines show comparisons to applicable published simulation results. We note that the right column is plotted from 0 to 2 instead of 0 to 3 in $r/R\sbr{V}$ to highlight the non-zero portion of the plot.}
    \label{fig:all_results}
\end{figure*}

In Fig.~\ref{fig:all_results}, we show the measured excess surface density profiles in the first column, and the corresponding three-dimensional void matter profiles in the second column for all five catalogues. The ESMD measurements are shown in black points, with the corresponding least-squares fits from the five-parameter HSW model in the black solid line and the void-galaxy correlation in the red dashed line. The blue lines represent 200 random samples of the MCMC posterior of the HSW fit to show the range of acceptable model fits given the priors discussed in Sect.~\ref{sc:model_fit}. The error bars on the data points come from the diagonal of the covariance matrix in Eq.~\eqref{eq:cov_stack} added with the covariance matrix measured from the signal around random lenses. The orange dot-dashed lines indicate published simulation results that roughly correspond to our catalogues. These will be discussed in Sect.~\ref{sc:discussion}. We additionally find that the cross-component of shear, $\gamma_\times$, is consistent with zero for all five catalogues and therefore do not plot them.

\begin{figure}
    \centering
    \includegraphics[width=\columnwidth]{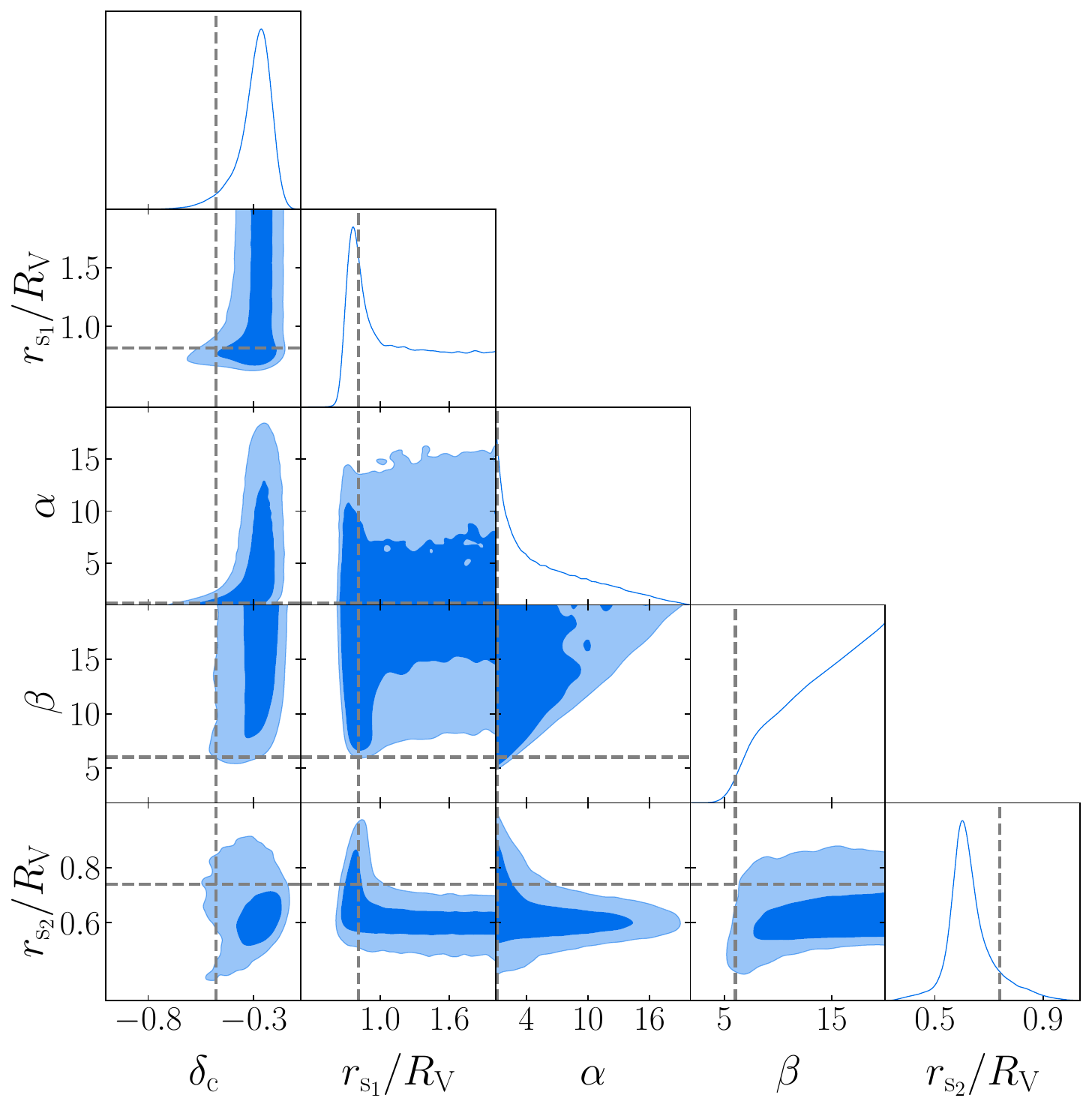}
    \caption{Contour plot of the five-parameter HSW model fit to the Full void catalogue. The least-squares fit is shown in the intersection of the dashed lines. Dark shaded regions indicate 1$\sigma$ contours, while light shaded regions indicate 2$\sigma$ contours.} 
    \label{fig:full_corner}
\end{figure}

The contour plot for the five free parameters of the modified HSW fit is shown, for the Full sample, in Fig.~\ref{fig:full_corner}. We note that the best-fit parameters, represented by the grey dashed lines, are located significantly far away from the mean of the marginalized posteriors in some cases. This is further explored in Appendix~\ref{sc:HSW} and may be due to the union of two distinct posterior regions of the five-parameter HSW model that produce similar quality fits to the data. We also find that this behaviour is not seen in four-parameter HSW model fits to our data. Fit statistics and parameter values are compiled for all five void catalogues in Table~\ref{tab:fit_stats}. For the parameter values, we report both the mean and 1$\sigma$ spread of the posterior, as well as the best-fit value in parentheses. Due to the complex nature of the posterior observed in Fig.~\ref{fig:full_corner} (e.g.\ the marginalized $\alpha$ PDF), in cases where the posterior is significantly skewed, we report the 95 per cent upper or 5 per cent lower limit.  For the significance measure of each profile, we report the area under the tail of the $\chi^2$ distribution used in Wilks' Theorem as described in Sect.~\ref{sc:model_fit} converted to the equivalent area under both tails of a Gaussian distribution, i.e.\ in units of $\sigma$, for convenience in interpretation.

From the $\Delta\Sigma$ plots of all 5 catalogues, we can clearly observe underdense structure, with minima at just short of the void radius, in agreement with the void lensing profiles of other works. The HSW fits from the MCMC chain visually trace a scatter in the ESMD profile predicted by the error bars, short of the last two or three bins. Fitting these last bins results in $\Delta \Sigma$ profiles that trend upwards at large scales, which result in model fits where $\alpha\approx\beta$. As discussed in Sect.~\ref{sc:model_fit}, these model fits produce 3D density profiles that are near-constant and non-zero at large scales, which are unphysical. We therefore have chosen a set of priors that reject these types of fits, which has resulted in $\Delta\Sigma$ profiles that asymptotically approach 0 at large scales. This is with the exception of the Small voids, whose $\Delta \Sigma$ measurement turns positive at comparatively much smaller radial bins, avoiding this issue. However, looking at the $\chi^2$ values, these restricted fits are still acceptable for all catalogues except for LOWZ, which is marginally rejected ($p\lessapprox0.05$). However, this marginal rejection still holds even when the prior bounds are relaxed. For the LOWZ profile in particular, the least-squares fit pushes for abnormally large values of $\beta$, where we report the best fit here as the maximal value of the prior bound set on $\beta$ for all catalogues (20). If this prior bound is relaxed, the LOWZ profile fits values of $\beta\approx250$ for a marginal change in $\chi^2$, which still results in marginally rejected fits. This indicates that the LOWZ catalogue does not place useful constraints on $\beta$.

\begin{figure}
    \centering
    \includegraphics[width=\columnwidth]{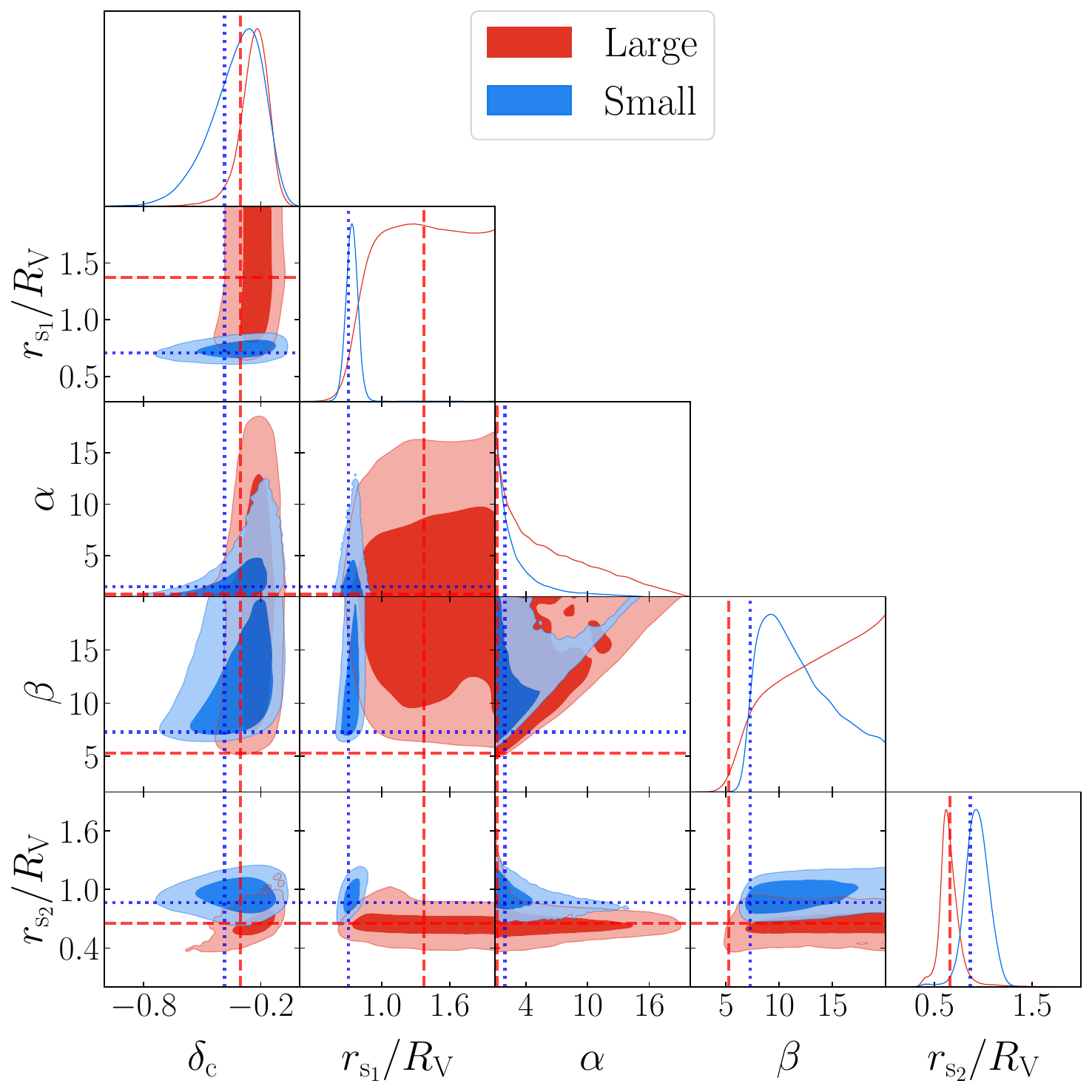}
    \caption{Corner plot of the HSW MCMC analysis for both the Large and Small void catalogues.}
    \label{fig:radsplit_corner}
\end{figure}

Comparing the profiles of the Small and Large samples, we can observe that smaller voids and larger voids seem to have substantially different $\Delta \Sigma$ profiles, with Small voids crossing zero just past the void radius, while Large voids cross zero at around 2.5 times the void radius. This is in line with findings that smaller voids tend to have more pronounced overdense walls than larger voids \citep{Hamaus2014,Nadathur2015}.  As these profiles represent the average of a stack of voids, this supports the notion that smaller voids are more likely than large voids to be ``void-in-clouds,'' where the smaller-scale underdensity may be situated in a larger-scale overdense environment. Looking at the corner plot in Fig.~\ref{fig:radsplit_corner}, we can see that these differences in the model fits may not be as substantial, and may be consistent with each other at the 1 to 2$\sigma$ level, with any difference being driven primarily by $r_\mathrm{s_1}$, which affects the zero-crossing scale, and $r_\mathrm{s_2}$, which primarily affects the large-scale behaviour of the voids. This consistency is driven mainly by the uncertainty in the Large model fits. However, if we subtract the void lensing measurements from each other directly, we find that the void lensing profiles of the Small and Large catalogues differ significantly at the 2.3$\sigma$ level.

To evaluate the linear biasing model, we can compare the predictions (red dashed curves) to the data. We note that the void-galaxy cross-correlation measurements that went into these models were measured to a maximum radius of 118 $h^{-1}\, \mathrm{Mpc}$ and as such will not extend across all radial bins of all catalogues. We find good visual agreement between the best-fitting void-galaxy model and the data for the Full and CMASS catalogues, which is supported by the $\chi^2\sbr{Vg}$ values reported in Table~\ref{tab:fit_stats}. Additionally, from the table, the void-galaxy fits to the Small catalogue, whose fits visually appear in the left column of Fig.~\ref{fig:all_results} of similar quality to perhaps the LOWZ fits, are actually a good fit to the data. The fits of the void-galaxy cross-correlation model to the Large and LOWZ catalogues are marginal, with $p$-values of 0.07 and 0.06, respectively. This may be because the model does not span the full extent of the data. To test, we can extrapolate the void-galaxy $\Delta\Sigma$ model by setting all values at radii larger than $118\, h^{-1}\,\mathrm{Mpc}$ to zero. We then find that the $p$-values for the Large and LOWZ catalogues become 0.13 ($\chi^2/\mathrm{dof}$=18.7/13) and 0.04 ($\chi^2/\mathrm{dof}$=22.9/13), respectively. While the extrapolated LOWZ catalogue produces a slightly poorer fit, the extrapolated fit has a $p$-value of the same order-of-magnitude as the HSW fit. In summary, these results suggest that the linear biasing model is not rejected by these measurements.

For the 3D overdensity plots in the second column of Fig.~\ref{fig:all_results}, we can compare the HSW fits and their scatter, extracted from the chains, with the void-galaxy tracer profile divided by the best-fitting $b\sbr{Vg}$ parameter. We note that for the Full and CMASS catalogues, the best-fitting matter profiles trace the void-galaxy profile and the general scatter fairly closely, except at small radii. There, the best fit (black line) has a linear slope on small scales while the chain samples tend towards constant values at small scales. This small-scale discrepancy is due to behaviour seen in the corner plot of Fig.~\ref{fig:full_corner}, which exhibits similar trends as in the CMASS catalogue. The resulting posterior contours are fairly complex, which results in the best fits for some parameters, notably $\delta\sbr{c}$, $\beta$, and $r_\mathrm{s_2}/R\sbr{V}$, represented by the grey lines, being found away from the mean of the posteriors. We do however find that, when compared visually, the best fitting models are in better agreement with the void-galaxy profiles than a model constructed from the mean of the posterior. This is why we additionally include the parameter values of the least-squares fit in Table~\ref{tab:fit_stats} in parentheses. We examine the properties of this complex posterior, including its mean, and how it may be due to the union of two distinct parameter regions of the five-parameter HSW model in Appendix~\ref{sc:HSW}.

The 3D matter density profiles of the Full, CMASS, and Large samples also indicate that these voids do not have a significant overdense wall, which is consistent with the void-galaxy profiles. This follows expectations from prior void analyses in simulations where larger voids tend to be compensated or undercompensated and thereby have less pronounced walls, as well as the use of the circumcentre definition for the void centres, which primarily presents with dampened walls in the density profiles in comparison to void profiles with the barycentre definition \citet{Nadathur2015b}. This is due to the minimum density centre not necessarily being located at the centre of the void, so the walls of the void become averaged out in the stacked profile. The LOWZ sample best fit HSW profile stands out with a comparably sharply peaked overdense wall, something that is not observed in the void-galaxy profile. This feature may be a result of the aforementioned weak $\beta$ constraints offered by this measurement. A similar overdense feature is prominent within the matter profile of the Small void catalogue as well, which is expected given that smaller voids tend to be overcompensated or ``void-in-clouds'', but the corresponding void-galaxy profile exhibits a weaker overdense feature. Although, as seen in Table~\ref{tab:fit_stats}, the void-galaxy model is still an acceptable fit to the Small catalogue.

\subsection{Integrated Density Contrast}
\label{sc:Results_integ_Density}
In order to assess the depth of the matter underdensities measured with this weak lensing analysis, in principle one can simply read off the best-fit values of the $\delta\sbr{c}$ parameter of the HSW model, as this represents the value of the profile at the center. However, this value represents an extrapolation of the HSW model fit to $R=0$. Since the innermost bins are the noisiest because they are based on the fewest galaxy shapes, $\delta\sbr{c}$ will not be well-constrained by the data. Instead, to get a more robust measure of the depth of the void, we can calculate the integrated density contrast, $\Delta$, defined as the volume-normalized integral to a radius $r$ of the HSW model in Eq.~\eqref{eq:B+15}
\begin{equation}
    \Delta\left(r\right)=\frac{3}{r^3}\int_0^r\delta\left(y\right)\, y^2\,dy\,.
\end{equation}
Then, choosing different values for the integrated radius can provide details about both the depth of the void stacks and how well each void stack is compensated (i.e.\ how closely the matter in the overdense wall can re-fill the void). To this end, we present the integrated density contrasts of the best-fit HSW models at three different values: $0.5 R\sbr{V}$, $R\sbr{V}$, and $3R\sbr{V}$ in Table~\ref{tab:Delta}.
\begin{table}
    \centering    
    \caption{Integrated density contrasts computed using the best-fit HSW profile shown in Fig.~\ref{fig:all_results} at three different radius values. Error bars represent the standard deviation in the MCMC samples.}
    \begin{tabular}{c|c|c|c}
        \hline
         & $\Delta\left(r=0.5R\sbr{V}\right)$ & $\Delta\left(r=R\sbr{V}\right)$ & $\Delta\left(r=3R\sbr{V}\right)$\\
         \hline
        Full & $-0.28\pm0.04$ & $-0.055\pm0.011$ & $\phantom{-}0.0003\pm0.0006$\\
        LOWZ & $-0.26\pm0.06$ & $-0.037\pm0.016$ & $-0.0010\pm0.0008$\\
        CMASS & $-0.24\pm0.07$ & $-0.063\pm0.017$ & $\phantom{-}0.0003\pm0.0008$\\
        Small & $-0.27\pm 0.07$ & $-0.001\pm0.025$ & $\phantom{-}0.0077\pm0.0023$\\
        Large & $-0.22\pm0.06$ & $-0.072\pm0.015$ & $-0.0025\pm0.0007$\\
    \end{tabular}
    \label{tab:Delta}
\end{table}
\begin{figure}
    \centering
    \includegraphics[width=\columnwidth]{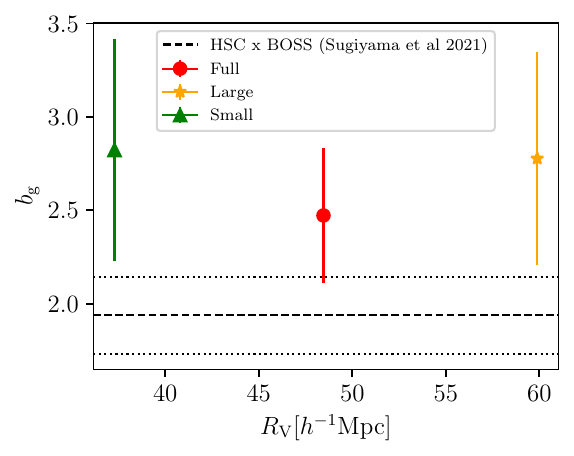}
    \caption{Galaxy bias estimates as a function of void size from comparing the void-galaxy cross-correlation to the void lensing measurements, as well as the large-scale galaxy bias measurement from \citet{Sugiyama21} for comparison. The dotted lines represent the 1$\sigma$ error on the averaged \citet{Sugiyama21} galaxy bias, represented by the dashed line. We note that the Full measurement does not constitute an independent measurement, but it is intended to show the full constraining power in this analysis.}
    \label{fig:galbias}
\end{figure}
We find that, unlike the distribution of $\delta\sbr{c}$, the distribution of $\Delta$ from the MCMC samples at all three radius values are Gaussian and centred near the value at the best fit. Therefore, we can use the standard deviation in the MCMC samples as the error bars for the $\Delta$ measurements. 

We observe that the integrated density contrasts out to half of the void radius are all consistent with each other for all 5 catalogues. This is somewhat expected for the Full, LOWZ, and CMASS catalogues as void depths are not expected to vary significantly over this range of redshift. What is of note is that the Small and Large catalogues are consistent with each other at this radius, which could be due to the lack of constraining power of the measurement as seen in Fig.~\ref{fig:all_results}. This is somewhat unexpected, as typically smaller \texttt{Revolver} voids tend to exhibit shallower central underdensities than larger voids \citep{Nadathur2015b}.

Looking to integrated density contrasts at the void radius, the Small and Large catalogues have a significant difference, with the Small catalogue $\Delta$ consistent with being compensated ($\Delta = 0$) at this scale, while the Large catalogue features the deepest density contrast of the five catalogues. This, coupled with the consistency at $0.5R\sbr{V}$ indicate that the differences in the density distributions comes largely from the placement and slope of the void boundary relative to the central underdensity. The Small catalogue's density profile, seen in Fig.~\ref{fig:all_results}, features a relatively close zero-crossing scale, with a prominent overdense wall, so it is reasonable that the integrated density returns to zero at the smallest radii, while the Large profile is the only one across all five profiles to not have crossed zero by $R\sbr{V}$, and as such is the deepest integrated density of the five.

The final column of Table~\ref{tab:Delta} provides information on the overall compensation of the different void catalogues. At a distance of $3R\sbr{V}$, one would expect to start probing the large-scale effect of the void on its surroundings. An overcompensated void would have a positive integrated density on these scales, while an undercompensated void would have a negative integrated density, indicating a much larger void-like environment. The Full, LOWZ, and CMASS are consistent with a compensated void, in agreement with the results of the simulated LOWZ voids of \citet{Nadathur2015}. The Small voids, which were consistent with compensated at $R\sbr{V}$, are now slightly overcompensated, in agreement with the relationship shown in Fig.~6 of \citet{Nadathur2015}. We additionally find the Large catalogue remains undercompensated at this scale. 

\subsection{Galaxy Bias in Underdense Environments}
\label{sc:Results_Galbias}
One of the primary purposes of measuring the galaxy bias in this data set was to test the findings of \citet{Pollina17}, who found from simulated voids that $b\sbr{Vg}$ depends on void size. One of the primary results of that work is that galaxies within smaller voids exhibit $b\sbr{Vg} > b\sbr{g}$ by roughly 17 per cent, but $b\sbr{Vg}$ asymptotically approaches $b\sbr{g}$, the large-scale galaxy bias, for the largest voids. In Fig.~\ref{fig:galbias}, we plot the galaxy bias constraints from our Small, Large, and Full catalogues, noting that since Full is the combination of the Small and Large void catalogues, it is not an independent measurement. Instead, it is meant to show the tightest possible constraint we can make with this data. We can compare our bias measurements to the results of \citet{Sugiyama21}, who constrained $b\sbr{g}$ for LOWZ and CMASS independently, using a combination of LOWZ, CMASS and Hyper Suprime Cam (HSC) data to jointly measure $\Delta\Sigma$ and projected galaxy clustering. We find that all three points are consistent with the averaged large-scale bias from \citet{Sugiyama21}, though notably for all measurements, $b\sbr{Vg}>b\sbr{g}$, and the error bars are too large compared to the fractional change observed in \citet{Pollina17} to make definitive claims regarding any possible trend in the results. We do note, however, that the consistently larger $b\sbr{Vg}$ measurement -- compared to $b\sbr{g}$ -- is consistent with the predictions of \citet{Nadathur2019b}, who find a discrepancy between the void-galaxy measurement and the void density profile of simulated voids as a function of the distance from the centre of the void of up to 25 per cent when using the $b\sbr{g}$ value in the linear bias model. This discrepancy would show up as an inflated value of $b\sbr{Vg}$ compared to $b\sbr{g}$. Comparing our result to \citet{Sugiyama21}, we find $b\sbr{Vg}/b\sbr{g} = 1.36\pm0.27$, consistent with a higher value of $b\sbr{Vg}$, but with uncertainties sufficiently large that we cannot exclude equality. 

Given that the innermost and therefore least dense region of a void is difficult to reliably estimate the galaxy density of, and is dependent on the choice of void centre, we then question whether these innermost bins are suitable to test the linear bias model. We attempt to exclude bins at less than $0.5\,R/R\sbr{V}$ from our fits to test. However, we find that doing so increases the values of the best-fit galaxy biases to scatter around a value of three in all cases while the error in the fits grow significantly, in some cases quadrupling in size. However, the larger error is expected, as stacked circumcentre voids typically do not have strong overdense walls, and thus there is not much signal to fit to once the central underdensity is excluded. We additionally find good visual agreement with the void-galaxy profile fit to the first three bins of the $\Delta\Sigma$ measurement. Therefore we do not wish to exclude these bins from the fits in order to get a realistic estimate of the error bars to determine how well the galaxy bias can be measured in data. We instead note that the $-1$ minimum value the void-galaxy profile takes at the centre around circumcentre voids represents the physical lower bound of possible underdensities. Therefore, this analysis still demonstrates realistic constraints of the spread of galaxy bias values, which are still too large to determine trends with void size.


\section{Discussion}
\label{sc:discussion_header}
\subsection{Comparison with Previous Results}
\label{sc:discussion}
There have been few detections of the weak lensing signal from cosmic voids. The result here, a $6.2 \sigma$ detection from 2975 voids, is only the third measurement of void-galaxy lensing using voids identified in spectroscopic tracers, and the second for spectroscopic ZOBOV-based voids. One goal of this analysis is to justify efforts in developing methods to interpret the resulting signal, whether it be for the purposes of cosmological parameter estimation or for understanding the formation and evolution of voids themselves, by showing that this signal is already detectable at high significance levels using currently available data. Given that void statistics, including the void matter profile, are sensitive to differences in void finders, survey geometry, and tracer properties \citep{Nadathur2015b,Massara2022}, it remains difficult to directly compare results from different works. As such, the most direct comparison we can make with this work is to the results of \citet{Melchior2014}, as they also used a ZOBOV-based void finder on a spectroscopic tracer sample from the 7th data release of SDSS. However, the 7th data release contains a catalogue of main sample galaxies, as well as a separate catalogue of LRGs. Using the model of \citet{Lavaux2012}, their most significant measurement combining both catalogues was 2.4$\sigma$, while the measurement using only the LRG catalogue was $1.3\sigma$. The second spectroscopic void lensing result comes from \citet{Clampitt15}, however they use a void finder that identifies voids in projected slices which may result in significantly different void profile properties. It is therefore not clear how to directly compare our 3D void lensing profiles to their void lensing profiles. 

We can compare the detection significance with that of other works as well, but we must note the caveats surrounding the methodologies. As an example, \citet{Fang2019} used the VIDE void finder on Year 1 of the DES survey, which is an imaging survey with photometric redshifts. For their 3D ZOBOV voids, they find a signal-to-noise ratio (SNR) of 14.0, using the equation from \citet{Becker16}:
\begin{equation}
\label{eq:sn}
    \mathrm{S/N}=\frac{\sum_{i,j}\Delta\Sigma_i^\mathrm{data}C_{ij}^{-1}\Delta\Sigma_j^\mathrm{model}}{\sqrt{\sum_{i,j}\Delta\Sigma_i^\mathrm{model}C_{ij}^{-1}\Delta\Sigma_j^\mathrm{model}}}\,.
\end{equation}
Using our measurement, we find a SNR of 5.52 for the Full sample, with a maximum of 6.42 for the Small sample, with the increase due to the more pronounced overdense wall present in smaller voids. While our detection is formally less significant than that of \citet{Fang2019}, it should be noted that, as demonstrated in simulations in that paper, 3D voids found from tracers with photometric redshifts experience a selection bias such that the resultant catalogue will be preferentially oriented along the line of sight. As shown in their Fig.~3, this results in a significant anisotropic void-galaxy cross correlation with the line of sight axis being 8 to 10 times the length of the other axis. Consequently, the projected density, to which weak lensing is sensitive, will have a much deeper underdensity than for voids found using spectroscopic tracers, providing a much stronger void lensing signal but at the cost of breaking assumptions of isotropy in void models. As pointed out in their work, this results in their HSW fits to the DES Y1 void lensing measurements fitting to the boundary value of $\delta\sbr{c}=-1$ in some cases.

There are two simulated void matter density profiles to which we can directly compare our results. They are both shown in Fig.~\ref{fig:all_results} as orange lines. The first is the LOWZ void density profile from Appendix A of \citet{Nadathur2015} chosen to match the average radius of our void lensing profile, and the second is the void profile from in Fig.~1 of \citet{Nadathur2019b}, although it should be noted that this profile is constructed of simulated CMASS voids, and includes all voids where $R\sbr{V}>43\,h^{-1}\,\mathrm{Mpc}$, while our Large catalogue includes LOWZ and CMASS voids, with a radius cut at around $48 \,h^{-1}\,\mathrm{Mpc}$. Both of these profiles come from simulated catalogues of circumcentre voids. We find that compared to these simulations, our 3D void density profiles are significantly shallower both within the void at small scales and at the outside wall at large scales. This results in the simulated profiles exhibiting deeper $\Delta \Sigma$ profiles than our measurements. However, when treating these simulated profiles as fixed models and comparing them to the data, we find that the simulated matter profile of Large voids \citep{Nadathur2019b} is an acceptable fit to the data, with $\chi^2/\mathrm{dof}$ of 19.4/14. As expected from a visual comparison, we find that the simulated profile of LOWZ voids \citep{Nadathur2015} is a poor fit to the data, with $\chi^2/\mathrm{dof}$ of $38.6/15$ ($p=7.4\times10^{-4}$). The increased depth of the simulated voids, particularly in the case of LOWZ, is an interesting result that may warrant further investigation, as it could be due to differences in the void selection, untested systematics in the void lensing methodology, or more physical in nature, such as the effect of neutrinos which would free stream into the void centre.

\subsection{Future Work}
\label{sc:Future}
The significance of this detection represents a promising step for the future of weak lensing around voids as a viable probe. It shows that currently available ground-based survey data is of sufficient quantity and quality to detect this weak effect. With the vastly improved data quality expected from the rapidly approaching advent of Stage-IV surveys like \textit{Euclid}, we expect to detect this signal with even greater precision (Martin et al., in prep). We also expect the detection from UNIONS to become stronger as more data becomes available from the survey. In particular, the inclusion of photometric redshift information would allow us to create narrower source bins in redshift and more accurately isolate and remove sources in front of our lenses, which act as a contaminating factor. Therefore, we would expect the constraints to improve significantly when the full photometric UNIONS catalogue is available. Furthermore, current spectroscopic surveys, such as DESI \citep{DESI2016}, will improve the coverage and quality of void catalogues in the NGC, and their expected overlap with UNIONS will provide another promising dataset with which to study void lensing.

This then raises the question of what additional analysis is needed to extract useful information from these void lensing profiles. One of the large uncertainties in extracting information from void statistics comes from the complex influence that the specific void finder has on a given statistic. As such, it may become necessary to simulate the specific void setup used in the data to properly test common assumptions and systematics, particularly for circumcentre voids, which are not as well studied as barycentre voids. For extracting the information itself, an emulator-based method may be ideal, as consistently running the same void finder across different cosmologies in a training set will allow an emulator model to learn how the specific void lensing profile changes with cosmology without the need to model out the relationship between the void finder and the void statistic. This is a procedure in development with void-galaxy cross-correlations and is seeing success \citep{Fraser2024}. Use of these emulators, depending on the variable parameters in the training simulation suite, should allow for constraints on cosmology, neutrino mass, and modified gravity.

\section{Conclusions}
\label{sc:conc}
This work represents a significant development in the detection of the void lensing signal from spectroscopic voids and the quality of available data for measuring it. To make this measurement, we used the substantial overlap of the UNIONS weak lensing survey with voids found in the foreground spectroscopic BOSS LOWZ and CMASS surveys. We also demonstrated that the use of mocks for selecting random lenses is necessary for voids to account for selection effects due to the survey boundaries. Additionally, we adapted and applied the standard Gaussian component of covariances of 2-point statistics to void lensing for the first time. By fitting the analytic matter density profile of \citet{Hamaus2014} with the modification of \citet{Barreira2015}, we determined the significance of our measurement to be 6.2$\sigma$, the most significant detection of 3D spectroscopic voids to date. We also compared our void lensing measurement to an adapted form of the void-galaxy cross correlation measurement from \citet{Woodfinden2022} in order to obtain a measure of the galaxy bias relative to underdense environments, at a value of $b\sbr{g}=2.45\pm0.36$ from the Full catalogue, consistent with the averaged large-scale galaxy bias value of \citet{Sugiyama21}. We find good agreement with a scale-independent linear bias model, but we do require more statistical precision in order to evaluate the findings of \citet{Pollina17} and determine the trend of $b\sbr{Vg}$ with void size. Our galaxy bias findings are however consistent with the findings of \citet{Nadathur2019b}, who predict that the large-scale linear galaxy bias will underestimate the bias between the void-galaxy correlation and the void density profile. We believe that higher statistical precision can be achieved with the full tomographic sample of the UNIONS survey, as tomographic binning will allow for better control and removal of foreground sources that act as interlopers in our signal. With the advent of next generation surveys such as \textit{Euclid} and LSST \citep{LSST2019} promising wider and deeper catalogues usable for both void finding and weak lensing, the future of exploring the darkest parts of the Universe is very bright.

\section*{Acknowledgements}
We would like to thank the anonymous referee for their insightful comments.

HM and MJH acknowledge financial support from the Canadian Space Agency (Grant 23EXPROSS1) and NSERC. LVW acknowledges support from an NSERC Discovery grant. LB is supported by the PRIN 2022 project EMC2 - Euclid Mission Cluster Cosmology: unlock the full cosmological utility of the Euclid photometric cluster catalog (code no. J53D23001620006). HH is supported by a DFG Heisenberg grant (Hi 1495/5-1), the DFG Collaborative Research Center SFB1491, an ERC Consolidator Grant (No. 770935), and the DLR project 50QE2305. PB acknowledges financial support from the Canadian Space Agency (Grant 23EXPROSS1), the Waterloo Centre for Astrophysics and the NSERC Discovery Grants program. FHP acknowledges support from CNES. MM is funded by the European Union (ERC, RELiCS, project number 101116027). Views and opinions expressed are however those of the author(s) only and do not necessarily reflect those of the European Union or the European Research Council Executive Agency. Neither the European Union nor the granting authority can be held responsible for them.

We are honoured and grateful for the opportunity of observing the Universe from Maunakea and Haleakala, which both have cultural, historical and natural significance in Hawai’i. This work is based on data obtained as part of the Canada-France Imaging Survey, a CFHT large program of the National Research Council of Canada and the French Centre National de la Recherche Scientifique. Based on observations obtained with MegaPrime/MegaCam, a joint project of CFHT and CEA Saclay, at the Canada-France-Hawaii Telescope (CFHT) which is operated by the National Research Council (NRC) of Canada, the Institut National des Science de l’Univers (INSU) of the Centre National de la Recherche Scientifique (CNRS) of France, and the University of Hawaii. This research used the facilities of the Canadian Astronomy Data Centre operated by the National Research Council of Canada with the support of the Canadian Space Agency. This research is based in part on data collected at Subaru Telescope, which is operated by the National Astronomical Observatory of Japan. Pan-STARRS is a project of the Institute for Astronomy of the University of Hawai’i, and is supported by the NASA SSO Near Earth Observation Program under grants 80NSSC18K0971, NNX14AM74G, NNX12AR65G, NNX13AQ47G, NNX08AR22G, 80NSSC21K1572 and by the State of Hawai’i.

This work was supported in part by the Canadian Advanced Network for Astronomical Research (CANFAR) and Compute Canada facilities.

Funding for the Sloan Digital Sky Survey has been provided by the Alfred P. Sloan Foundation, the Heising-Simons Foundation, the National Science Foundation, and the Participating Institutions. SDSS acknowledges support and resources from the Center for High-Performance Computing at the University of Utah. The SDSS web site is www.sdss.org.

\section*{Data Availability}
The void catalogue is available upon request from the authors.
A subset of the raw data underlying the source catalogue used in this article is publicly available via the Canadian Astronomical Data Center\footnote{\url{http://www.cadc-ccda.hia-iha.nrc-cnrc.gc.ca/en/megapipe/}}. The raw  and processed UNIONS data are currently available to members of the Canadian, French, Japanese and Pan-STARRS communities. All UNIONS data will be publicly available to the international community at the end of the proprietary period.


\bibliographystyle{mnras}
\bibliography{example} 


\appendix

\section{Comparing Covariance Terms}
\label{sc:appendix_cov}
As discussed in Sect.~\ref{sc:cov}, the full analytic covariance matrix without simplifications can be written as
\begin{multline}
\label{eq:cov_appendix}
    \mathrm{Cov}\left[\Delta\Sigma\left(\theta_1\right),\Delta\Sigma\left(\theta_2\right)\right]=\frac{1}{4\pi f\sbr{sky}\sigcinv^2}\int \frac{d\ell}{2\pi}\widehat{J_2\left(\ell\theta\sbr{1}\right)}\widehat{J_2\left(\ell\theta\sbr{2}\right)} \\\times\left[\frac{\ckk\left(\ell\right)}{n\sbr{V}}+C\sbr{VV}\left(\ell\right)\ckk\left(\ell\right)+C\sbr{VV}\left(\ell\right)\frac{\sigma^2_\epsilon}{n\sbr{g}}+\cvk^2\left(\ell\right)\right]\\+\delta^K_{\theta_1,\theta_2}\sigma^2\sbr{SN,\Delta\Sigma}\,,
\end{multline}
where we have expanded the binomial multiplication seen in Eq.~\eqref{eq:krause_cov} and converted from the units of shear to the units of $\Delta \Sigma$. We now want to compare the contributions from each of the terms in square brackets to the total covariance. In order to obtain analytic formulas for each of the angular power spectra, $C\sbr{AB}\left(\ell\right)$, here, we use Limber's approximation \citep{Limber53,Loverde08} to write the general formula for any tracers A,B:
\begin{equation}
\label{eq:CAB}
    C\sbr{AB}(\ell)=\int d\chi\frac{W\sbr{A}\left(\chi\right)W\sbr{B}\left(\chi\right)}{\chi^2}P\sbr{AB}\left(k=\frac{\ell+1/2}{\chi}\right)\,,
\end{equation}
where $W\sbr{A,B}$ are tracer-dependent weighting functions. For example, if we write
\begin{equation}
\label{eq:W_k}
W_\kappa\left(\chi\right)=c\overline{\rho}\int_{z\left(\chi\right)}^\infty dz\sbrc{S} \frac{n\left(z\sbrc{S}\right)}{H(z\sbrc{S})\Sigma\sbr{c}\left(z\sbrc{S},\chi\right)}\,,
\end{equation}
and we note that $P_{\kappa\kappa}=P\sbr{mm}$ as lensing is sensitive to the underlying matter distribution, we end up with the definition of $\ckk$ that we used in Eq.~\eqref{eq:ckk}.

In order to define $W\sbr{V}$, we follow the definition used in \citet{Bonici22}:
\begin{equation}
\label{eq:W_V}
    W\sbr{V}\left(z\right)=\frac{H\left(z\right)}{c}n\sbr{V}\left(z\right)\,b\sbr{V}\,,
\end{equation}
where $b\sbr{V}$ is the void bias. Incorporating the void bias into the weighting function here allows us then to write the void-void power spectrum, $P\sbr{VV}$ like so:
\begin{equation}
\label{eq:pvv}
    P\sbr{VV}\left(k,z\right)=\left[1-S\sbr{N}\left(k\right)\right]P\sbr{mm}\left(k,z\right)\,,
\end{equation}
where $S\sbr{N}\left(k\right)$ is a low-$k$ pass filter to cut out small scale information. This is done to avoid the need to model the effects of the void exclusion principle, wherein two voids cannot overlap too strongly, else they are treated as a single void by a void finder. Since we are only interested in an order of magnitude estimate here to gauge the level of contribution these terms have to our covariance, modelling the small-scale information including the void exclusion principle here is outside of the scope of this work. The low-k pass filter is the smoothstep function:
\begin{equation}
    S_N\left(k\right)=
    \begin{cases}
        0&\textrm{if}\;\frac{k}{k_\mathrm{ref}}{\leq} a \\
        \left(\frac{k}{k\sbr{ref}}\right)^{N+1}\sum\limits_{n=0}^N\binom{2N+1}{N-n}\binom{N+n}{n}\left(-\frac{k}{k\sbr{ref}}\right)^n & \textrm{if}\; a{\leq}\frac{k}{k\sbr{ref}}{\leq} b\,, \\
        1 & \textrm{if}\;\frac{k}{k\sbr{ref}}{\geq} b
    \end{cases}
\end{equation}
with parameters set similar to those in \citet{Bonici22}: $a=1$, $b=1.8$, $N=3$, and the reference scale $k\sbr{ref}$ is determined to be the scale of the average void radius: $k\sbr{ref}=2\pi/R\sbr{V}$. 

\begin{figure}
    \centering
    \includegraphics[width=\columnwidth]{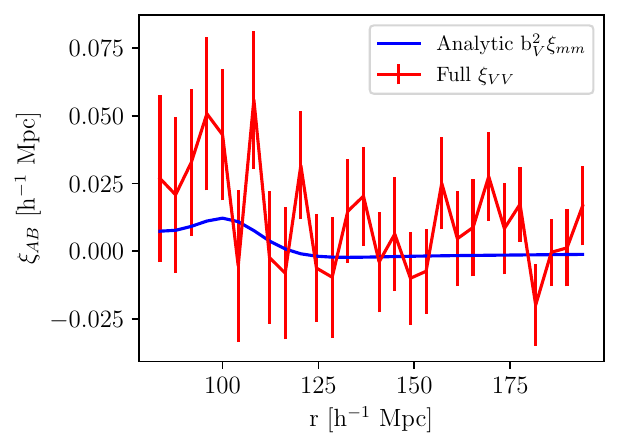}
    \caption{Measurement of the void bias from the BOSS void sample using the void-void correlation measurement from the Full catalogue and an analytical calculation of the matter-matter correlation function. The fit is performed on scales larger than two times the average void radius to remove the void exclusion effect from the analysis.}
    \label{fig:bv_fit}
\end{figure}

Therefore, in order to compute terms involving voids as tracers, we need an estimate of the void bias. We initially attempt to measure it from our data. If we measure the void-void clustering signal $\xi\sbr{VV}$, we can compute the matter-matter correlation function from the fiducial cosmology. Since we are cutting out the small scale information, these two quantities are related by the void bias factor:
\begin{equation}
    \xi\sbr{VV}\left(r\right)=b\sbr{V}^2\xi\sbr{mm}\left(r\right)
\end{equation}
We measure the void-void clustering signal using \texttt{pycorr} \citep{pycorr}, which uses jackknife to determine the error bars, then compute the matter-matter correlation function using \texttt{Colossus} \citep{Diemer18} and the fiducial cosmology. However, as seen in \citet{Clampitt16}, the void bias is only measurable in configuration space between scales corresponding to twice the average void radius (due to the void exclusion) and the baryon acoustic oscillation (BAO) peak, which affects the matter correlation function. Because of these effects, \citet{Clampitt16} measured their void bias from $2R\sbr{V}$ to $80 h^{-1}$Mpc. Unfortunately, because our voids are traced by LRGs, which are a sparse and highly biased tracer sample, our average (unweighted) void radius is around 40 $h^{-1}$Mpc. This means that we would have no region within which to safely determine the void bias. However, again, as we are only interested in an order of magnitude estimate here, we instead attempt to measure the void bias in the region 2 to $5 R\sbr{V}$. We then take the computed matter-matter correlation function, multiply it by a free parameter of $b\sbr{V}^2$, and then perform a least-squares fit to the void-void correlation measurement from \texttt{pycorr}. The results of this fit are shown in Fig.~\ref{fig:bv_fit}. We can observe that the large size of an average void in our catalogues means that the primary positive matter-matter correlation signal that we fit is the BAO peak. From studies that have attempted to measure the BAO signal from void clustering \citep{Liang2016,Kitaura2016b}, it was demonstrated that non-overlapping voids do not carry the BAO signal. Since ZOBOV voids are by construction non-overlapping, we do not expect the BAO peak to be included in our clustering profile. However, excluding the BAO peak results means that the primary matter-matter signal on these large scales is negative, resulting in unphysical negative fits to the square of the void bias. The same holds true even when fitting to smaller scales like $1.5R\sbr{V}$. We therefore come to the conclusion that we are unable to constrain a reasonable estimate of the void bias from void-void clustering. 

Instead, since we are interested in an order of magnitude estimate, we can use the results of \citet{Jamieson2019}, who measured the void bias in Separate Universe Simulations from \texttt{VIDE} voids. They measure a void bias at a redshift of 0.5 that ranges from roughly 1.5 at the smallest voids, to -1 at the largest, crossing 0 at roughly 60 $h^{-1}\, \mathrm{Mpc}$. These large-scale void bias measurements are of similar order as those of \citet{Hamaus2014b}, who used simulated voids with tracer densities matching SDSS DR7. As such, we assume a void bias value of 1.5 for Small, 0.5 for Full, LOWZ, and CMASS, and a value of -1 for Large. We assume a negative value for Large despite its average void radius coinciding with the $b\sbr{V}=0$ size as we find from the integrated density contrast that this catalogue is expected to be undercompensated. A bias of zero would indicate compensation. In contrast, we find that Full, LOWZ, and CMASS are compensated, indicating very small values of the void bias. We choose to set the void bias to 0.5 as a conservative estimate for this order of magnitude approximation for $\cvk$ and $C\sbr{VV}$. 

\begin{figure}
    \centering
    \includegraphics[width=\columnwidth]{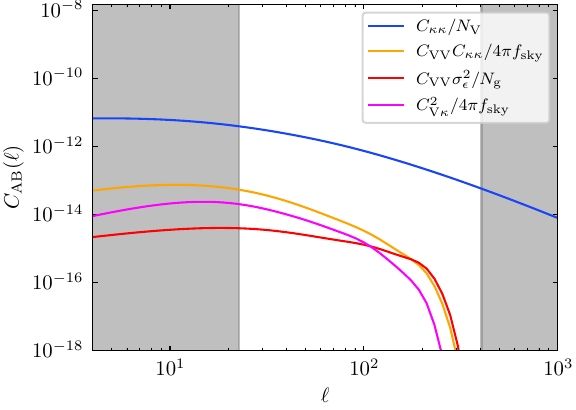}
    \caption{Comparison plot of the different expanded terms of the analytic covariance model for the Full void catalogue. The non-grey region includes all scales between the maxima of the Bessel window functions for the largest ($3R\sbr{V}$) and smallest bins used in the Full catalogue.}
    \label{fig:cov_term_compare}
\end{figure}

Using these values of the void bias, we can then compute $C\sbr{VV}$ and $\cvk$ using Eqs.~\eqref{eq:CAB}, \eqref{eq:W_k}, and \eqref{eq:W_V}, using \texttt{Colossus} \citep{Diemer18} to compute the linear matter power spectrum. Then, we can compute the four terms that exist inside of the square brackets of Eq.~\eqref{eq:cov_appendix}. We note that it is not clear what a reasonable estimate of the sky fraction would be, as we include voids whose centre exists outside of the overlap of the BOSS and UNIONS footprints. Given that the void binning is so large in this case, these voids will still have a significant number of lens-source pairs, but create a ``fuzzy'' edge around the intersection where void centres exist while sources do not. In initial testing, we found that the comparison to random covariance, which we used for validation in Fig~\ref{fig:covcompare} differed by about 10 per cent in all bins when using the strict intersection of the two surveys. For the purposes of these plots, we choose a sky fraction value that corresponds to the strict intersection, with the understanding that this will be a slight overestimation in the final covariance terms. This area problem is avoided in the final covariance model in Eqn.~\ref{eq:cov_stack} under the assumption that $\cvk$ and $C\sbr{VV}$ are negligible. We plot all four terms in Fig.~\ref{fig:cov_term_compare} over a range of scales, where we observe that the $\ckk/N\sbr{V}$ term is larger than all other terms on the plotted scales by more than an order of magnitude. This then provides reasonable justification to exclude the other three terms from the covariance formula.

\section{Investigating the Five-Parameter HSW Model Posterior}
\label{sc:HSW}
In this section we attempt to explore properties of the posterior of the HSW model with the B+15 modification (see Eq.~\ref{eq:B+15}) to void matter profiles of circumcentre voids. Generally, the four-parameter model of \citet{Hamaus2014} has been used in the literature (with the exception of modified gravity studies), which originally fixed $r_\mathrm{s_2} = R\sbr{V}$. The decision to use this modified model for our analysis was motivated by the findings of \citet{Nadathur2015}, who found qualitatively that the five-parameter model produces a better fit to circumcentre voids, while the HSW model was initially empirically determined using barycentre voids. This was further motivated by the particular results after fitting the four-parameter HSW model to our measurements that will be detailed later in this section.

As encountered in Fig.~\ref{fig:full_corner}, the resulting posterior of the five-parameter fit to the Full catalogue (with similar complexities present in all catalogues but Small) is fairly complex and non-Gaussian. It further appears that some parameters, such as $r_\mathrm{s_1}/R\sbr{V}$ and $\beta$ contain regions where these values are unconstrained, while others, such as $\delta\sbr{c}$ and $r_\mathrm{s_2}/R\sbr{V}$ are much more constrained and Gaussian-like. An additional peculiarity present in this corner plot is the location of the least-squares best fits is often two or more $\sigma$ away from the peak of the marginalized histograms. Since these posteriors were built from the steps of an ensemble of MCMC walkers, a natural question to ask is whether the chains have sufficiently converged. To validate this, we used the \texttt{dynesty} package \citep{Speagle2020}, which runs a nested sampler algorithm to estimate the posterior as opposed to a random-walk approach. This represents an independent constraint on the posterior which additionally has the advantage of a natural definition of a stopping criterion. Comparing the resulting contours, we found little change in the contour shape, indicating that our MCMC analysis was converged.
\begin{figure*}
    \centering
    \includegraphics[width=\textwidth]{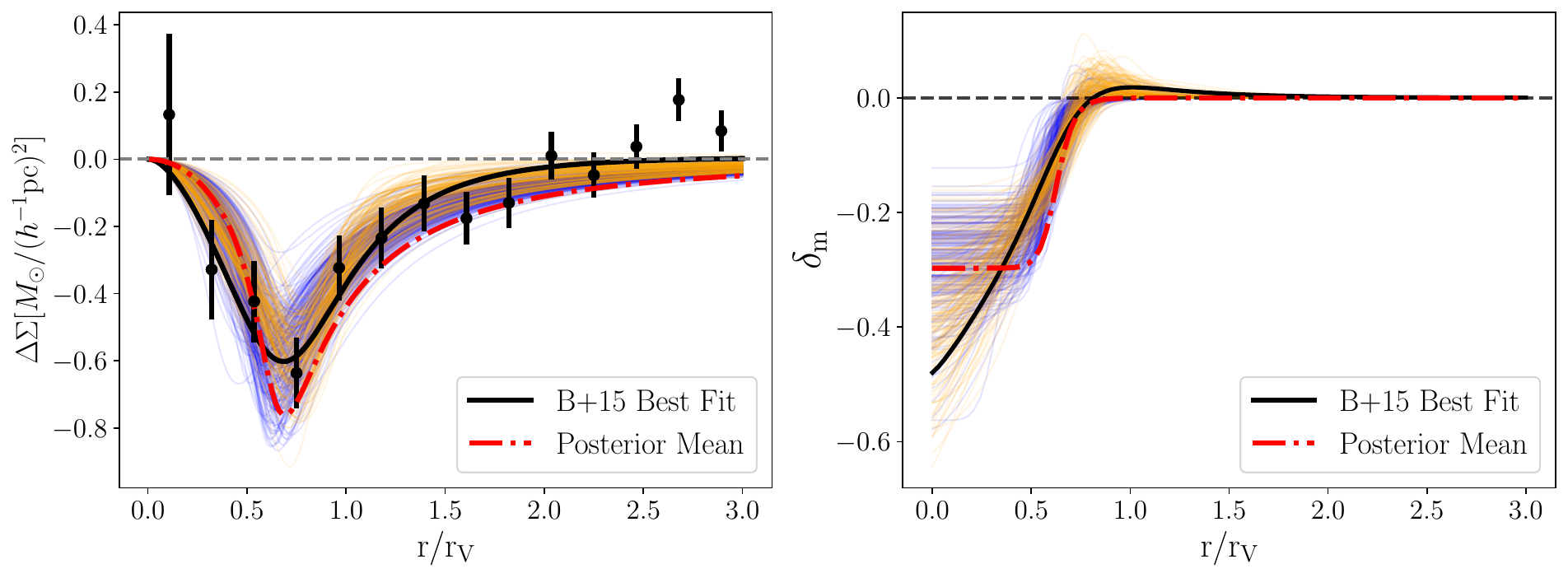}
    \caption{The $\Delta\Sigma$ and $\delta\sbr{m}$ plots of the Full sample as shown in Fig.~\ref{fig:all_results}. The orange lines represent 200 MCMC samples from the posterior region that includes the best-fit model, represented by the black line. The blue lines represent 200 MCMC samples from the posterior region where the $\delta_m(r/r_{\rm v})$ profile is step-function-like, which occurs when $r_\mathrm{s_2}$ is sufficiently smaller than $r_\mathrm{s_1}$. Additionally, we plot the model corresponding to the posterior mean in the red dot-dashed line, measured by taking the average of MCMC samples in each model parameter.}
    \label{fig:full_color}
\end{figure*}

A potential explanation for the behaviour of this posterior comes from studying the shape of the contour in the $r_\mathrm{s_1}/R\sbr{V}$--$r_\mathrm{s_2}/R\sbr{V}$ plane. This L-shaped contour seems to indicate that there are two distinct regions of parameter space that are fit here -- one with large values of $r_\mathrm{s_2}/R\sbr{V}$, and one with small values. The large-valued region, which includes the best fit, seems to provide a narrow, constrained region of the $r_\mathrm{s_1}$ parameter, while the small-valued region seems only able to place a lower bound, with a mostly uniform sampling density that appears to extend past the upper prior bound, if allowed. Looking at the $\delta\sbr{c}$--$r_\mathrm{s_1}/R\sbr{V}$ and the $\delta\sbr{c}$--$\alpha$ contours, we see similar contour shapes, denoting a distinction between high values of $\delta\sbr{c}$ and low values, where the high value regions indicate unconstrained regions of the other parameter, while low value regions contain the best fits and are more constrained. 

Examining the formula for the five-parameter model itself, we note an interesting relationship between the $r_\mathrm{s_1}$ and $r_\mathrm{s_2}$ parameters. If $r_\mathrm{s_2}$ is sufficiently smaller than $r_\mathrm{s_1}$, then the profile sharply transitions to 0 from the effect of the denominator at a scale before the profile crosses 0 from the effect of the numerator. The resulting 3D density profile appears more like a step-function, featuring a flat underdense region, a sharp near-vertical transition to 0, then a flat, 0 region on large scales. This step-function-like profile is relatively insensitive to changes in $\alpha$ and $\beta$, who normally are largely responsible for the slope from underdense to 0, and the slope from overdense to 0 in a typical void profile. Further, it is insensitive to changes in $r_\mathrm{s_1}$ above some sufficient lower bound, as changing where the profile crosses 0 after the denominator forces it to asymptotically approach 0 is immaterial to the overall profile's shape. Therefore, from this limiting behaviour of the formula itself, if there is a combination of values of $\delta\sbr{c}$ and $r_\mathrm{s_2}/R\sbr{V}$ that fit the data with $r_\mathrm{s_2}$ sufficiently smaller than $r_\mathrm{s_1}$, then the other three parameters are unconstrained.

However, it can be seen from the 3D density  plots of the least-squares best fit in Fig.~\ref{fig:all_results} that the best-fit profiles do not exhibit these step-function-like behaviours, and according to the corner plot for the Full fit, this best-fit lies within the smaller, more constrained region associated with larger values of $r_\mathrm{s_2}$. This indicates that there seem to be two distinct regions of parameter space fit by this posterior. One region where the fit is primarily determined by $\delta\sbr{c}$ and $r_\mathrm{s_2}$ and is unconstrained by the rest of the parameters, and one where the fit is determined by most if not all of the parameters. The typical difference in $\chi^2$ between these two regions is roughly 3, so both regions represent acceptably likely fits to the data. However, due to the fact that these corner plots are constructed from steps taken from the ensemble of walkers of the MCMC analysis, the former region is over-represented given the unconstrained nature of some of the parameters, and the fact that this unconstrained region is flat in likelihood space. This would explain why the best fitting model is located away from the mean of the marginalized histograms, particularly in $\delta\sbr{c}$ and $r_\mathrm{s_2}$. 

We additionally noted in Sect.~\ref{sc:results} that visually, in the 3D density profiles, the best fits did not seem fully represented by the scatter presented by the MCMC samples. Noting the two-region posterior, we expect the scatter to be more representative if we isolate the region that contains the best fit. In order to demonstrate this effect, we make a cut in $r_\mathrm{s_2}$ at a value of 0.73 to separate these two regions for the Full catalogue. The value of 0.73 was chosen to primarily exclude the horizontal region of the $r_\mathrm{s_1}$--$r_\mathrm{s_2}$ contour. We show the results of this cut in Fig.~\ref{fig:full_color}. The MCMC samples with $r_\mathrm{s_2}>0.73$ are coloured orange to represent the region of the posterior that includes the best fits. The blue samples represent the region whose fits are determined by only $\delta\sbr{c}$ and $r_\mathrm{s_2}$. For this plot, we used only the Full measurement, and plotted 200 samples from each posterior region. While difficult to tell from the $\Delta\Sigma$ plot, which indicates that both posterior regions fit the data well, we can clearly see in the 3D density plot that the scatter from the orange samples better represents the best fit line, while the scatter in the samples shown in blue features a steeper transition from underdense to overdense, more representative of the step-function-like posterior region.

We additionally plot the five-parameter HSW model corresponding to the mean of the projected 1D posteriors, taken as the average of the MCMC samples in each of the five parameters individually, in the red dot-dashed lines in Fig.~\ref{fig:full_color}. We can observe in the plot on the right that the mean of the posterior is representative of the step-function-like profiles shown by the subset of the MCMC samples in blue, indicating that the larger volume of the step-function-like posterior region is more prominently featured within the average. We do however observe that the posterior mean represents an extreme value of the blue scatter on mid-to-large scales in the left plot. This is likely due to the skewed nature of some of the marginalized histograms in the contour plot (notably $\alpha$). It is for these reasons we opted to plot the best fitting model as opposed to the posterior mean in Fig.~\ref{fig:all_results}.

The existence of these two posterior regions then raises the question of whether this is an effect of the use of \texttt{Revolver} voids or whether it is due to the increased flexibility offered by the five-parameter model from the B+15 modification over the four-parameter HSW model. It can be seen in the corner plot of Fig.~9 of \citet{Fang2019} that the four-parameter model used on \texttt{VIDE} void lensing signals does not produce the step-function-like parameter region, as $r_\mathrm{s_1}/R\sbr{V}<1$, which for the four-parameter model fits (where $r\sbr{s_2}/R\sbr{V}=1$) implies $r\sbr{s_1}<r\sbr{s_2}$. We see similar results in corner plots produced by fitting our measurements to the four-parameter model, indicating that this secondary parameter region is a result of the flexibility of the five-parameter model. Further, we find that the weak compensating walls that are a consequence of the circumcentre voids do contribute to how acceptable these types of step-function-like fits are. The Small catalogue, which has a significant overdense wall, does not have these step-function-like MCMC samples in the right-hand column of Fig.~\ref{fig:all_results}, and we find no similar complex contours in Fig.~\ref{fig:radsplit_corner}. We note that these types of profiles with almost no walls have not been directly observed in the literature for ZOBOV voids, save for the largest voids of a catalogue. In principle, this region could be excluded by the priors, potentially by a prior on the relationship between $r\sbr{s_1}$ and $r\sbr{s_2}$. However, we additionally observe from the differences in best-fit $r\sbr{s_1}$ and $r\sbr{s_2}$ values in Table~\ref{tab:fit_stats}, that the value at which $r\sbr{s_2}$ is sufficiently smaller than $r\sbr{s_1}$ is dependent on other properties, particularly in the case of Large, which exhibits the largest difference between the two. In order to construct a robust set of priors that can exclude this step-function-like region from the five-parameter fit, more investigation is needed. 

With this additional region of parameter space in mind, we also wanted to quantitatively evaluate the difference between our measurement as fit by the original four-parameter HSW model versus the fit provided by the B+15 modification. We can see from the best fit results shown in Table~\ref{tab:fit_stats} that profiles prefer values of $r_\mathrm{s_2}/R\sbr{V}$ that are slightly less than 1, with the exception of LOWZ and Small voids. Intuitively, this suggests that these measurements prefer the increased flexibility of the B+15 modification. Since these are nested models, we can test this by applying Wilk's Theorem to compare the significance in the best-fitting $\chi^2$ between the four- and five-parameter HSW models. From this, the only significantly improved fit comes from the Full measurement, at a marginal significance of $1.9\sigma$. For the other samples, the improvement is not significant, indicating that the additional degree of freedom does not provide a strong change in the quality of the fits. However, there is a caveat with the four-parameter model fits. With every catalogue, all four-parameter models have $\alpha$ at the prior boundary ($\alpha=1$), and upon relaxing this boundary, all fit to values $\alpha<1$, which produces voids with concave behaviour in the innermost region. Voids with these types of profiles have never been observed in simulations or in void measurements in the literature. It is therefore likely that the underlying cause is the inability of the data to constrain the innermost regions of the voids particularly well. This is likely due to the smallest bins in $\Delta\Sigma$ having the largest error bars, and therefore accepting a wide range of possible inner regions. This effect couples with the weak walls of circumcentre void stacks to produce weak constraints on $\alpha$. We observe this effect represented by extreme values of both $\alpha$ and $\delta\sbr{c}$ as seen in the four-parameter model and seemingly to a lesser extent (via the linear slopes) in the five-parameter modification. Because of the slightly better interpretations in the innermost regions, we opt to present the results of five-parameter HSW model over the four-parameter, but we note that both four- and five-parameter models can be used interchangeably for our results. 

\bsp	
\label{lastpage}
\end{document}